\newcounter{myctr}

\documentclass{ws-acs}

\usepackage{float}
\usepackage{mathrsfs} 
\usepackage{multirow} 
\usepackage{hhline} 
\usepackage{longtable}
\usepackage{algorithm}

\usepackage{algorithmic}

\usepackage{xspace}
\usepackage{subfigure}

\usepackage[compress]{cite}
\usepackage{xcolor}
\usepackage[verbose]{hyperref}
\hypersetup{colorlinks=false,allbordercolors=blue,pdfborderstyle={/S/U/W 1}}

\begin{document}

\makeatletter
\def\@biblabel#1{[#1]}
\makeatother

\markboth{Baby C. J. and Kamalika Bhattacharjee}{Hierarchical Clustering using Reversible Binary Cellular Automata for High-Dimensional Data}

%
\catchline{}{}{}{}{}
%

\title{Hierarchical Clustering using Reversible Binary Cellular Automata for High-Dimensional Data}

\author{Baby C. J.}

\address{Department of Computer Science and Engineering,\\ National Institute of Technology, \\
Tiruchirappalli, Tamilnadu - 620015 ,
India\\
\email{babycj1120@gmail.com}}

\author{Kamalika Bhattacharjee}

\address{Department of Computer Science and Engineering,\\ National Institute of Technology, \\
Tiruchirappalli, Tamilnadu - 620015 ,
India\\
\email{kamalika.it@gmail.com}}

\maketitle


\begin{abstract}
This work proposes a hierarchical clustering algorithm for high-dimensional datasets using the cyclic space of reversible finite cellular automata. In cellular automaton (CA) based clustering, if two objects belong to the same cycle, they are closely related and considered as part of the same cluster. However, if a high-dimensional dataset is clustered using the cycles of one CA, closely related objects may belong to different cycles. This paper identifies the relationship between objects in two different cycles based on the median of all elements in each cycle so that they can be grouped in the next stage. Further, to minimize the number of intermediate clusters which in turn reduces the computational cost, a rule selection strategy is taken to find the best rules based on information propagation and cycle structure. After encoding the dataset using frequency-based encoding such that the consecutive data elements maintain a minimum hamming distance in encoded form, our proposed clustering algorithm iterates over three stages to finally cluster the data elements into the desired number of clusters given by user. This algorithm can be applied to various fields, including healthcare, sports, chemical research, agriculture, etc. When verified over standard benchmark datasets with various performance metrics, our algorithm is at par with the existing algorithms with quadratic time complexity.
\end{abstract}


\section{Introduction}\label{sec1}

{C}{lustering} is a popular data processing technique that groups data based on some \emph{similarity metrics} without any supervision \cite{CL1, CL2, CL3}. Several clustering algorithms exists in the literature, like $K$-means \cite{hartigan1979algorithm, likas2003global}, Meanshift \cite{comaniciu2002mean}, BIRCH \cite{zhang1997birch}, DBSCAN \cite{ester1996density} and Hierarchical \cite{JSSv025i04}. However, in the last few years, reversible binary cellular automata  (CAs) based clustering algorithms have established reversible cellular automaton (CA) as a natural clustering technique \cite{manoranjan2023optimized} \cite{mukherjee2021clustering, mukherjee2021reversible,abhishek2023cellular}.

In a clustering algorithm, the data objects are divided into some groups (clusters) such that an object can belong to only one cluster. That means, a clustering algorithm is equivalent to a bijective function ${\mathcal{F} : \mathcal{D} \rightarrow \mathcal{D}}$ where $\mathcal{D}$ is the set of data objects. If the clustering algorithm is \emph{good}, it distributes the objects in such a way that the objects belonging to the same cluster are \emph{similar} or \emph{close} with respect to some metric, whereas, those belonging to different clusters are \emph{dissimilar} or \emph{distant}. So, in terms of $\mathcal{F}$, a pair of elements $x$ and $y$ are \emph{similar} if $x$ and $y$ are \emph{reachable} from each other, that is, $\mathcal{F}^{k_1} (x) = y$, $\mathcal{F}^{k_2} (y) = x$, for some $k_1, k_2 \in \mathbb{N}$, whereas, they are \emph{dissimilar} if there exists no $k \in \mathbb{N}$, such that, $\mathcal{F}^{k} (x) = y$ and vice versa. We can define an equivalence relation $\mathscr{R}$ to depict this \emph{similarity}: $x \mathscr{R} y$ holds if and only if, for some $k_1, k_2 \in \mathbb{N}$, $\mathcal{F}^{k_1} (x) = y, \mathcal{F}^{k_2} (y) = x$. This $\mathscr{R}$ creates several distinct partitions of $\mathcal{D}$ where each of these partitions represents an \emph{equivalent} class forming a unique cluster. Note that, the number of possible bijective functions for the dataset $\mathcal{D}$ is $(|\mathcal{D}|)!$, and for different choices of this function, data elements can be partitioned differently. Therefore, for any dataset, finding the appropriate clustering function $\mathcal{F}$ is a challenging task.

In the the case of reversible cellular automata, this issue is addressed by exploiting the global transition functions which are bijective. Here, the data elements are considered as configurations. So, for a reversible CA, the similarity metric is the \emph{reachability} of two configurations from each other -- only then they are part of the same cluster. That means each cycle is a cluster in a reversible CA. By proper selection, we can find the global transition function that distributes these configurations (data elements for our case) into meaningful clusters concerning that data set. However, a problem of the CA-based algorithms is, as the number of clusters is dependent upon the number of cycles of the CA, we may not directly get the desired number of clusters. For this reason, in Ref. \cite{mukherjee2021clustering, mukherjee2021reversible} an iterative hierarchical approach has been taken to further group the clusters generated in the previous level in the current level until the desired number of clusters is achieved.  Here, at every level, this grouping is done by using a new reversible CA rule. Our proposed algorithm can be applied to various fields, including healthcare, chemical research, agriculture, and more

However, this scheme has some limitations. The most important among them is, as clusters are equivalent to the cycles, the whole $n$-length configuration space of the $n$-cell CA has to be explored every time we use a CA, where $n$ is the size of the (encoded) data elements. That is, considering binary CAs, the complexity of $2^n$ limits our data element size, which in turn restricts the maximum number of features each data can have. For this reason, clustering high dimensional datasets using the schemes of Ref. \cite{mukherjee2021clustering, mukherjee2021reversible} are infeasible. Furthermore, out of the total number of reversible $2$-state $r$-radius CAs available for any $r$, particularly which CAs are best suitable for clustering has not been identified. So, one has to exhaustively run over the set of potential candidate rules and take whichever gives the best result. This approach is also not practical.

In this scenario, this work targets to address both of these issues. Like Ref. \cite{mukherjee2021clustering, mukherjee2021reversible}, we take only the one-dimensional reversible CAs with $2$ states per cell where each cell depends on itself and its two consecutive cells on both sides (that is, the radius is $2$) as neighborhood dependency under null boundary condition. However, the algorithms of these papers have exponential complexity over $n$ as the whole configuration space of $n$ cells needs to be explored. In this paper, to limit the complexity, we do the clustering in three stages. First, we split the (encoded) data frames into several partitions such that each partition is within our computational limit and apply the reversible CA rule in parallel to each of these splits to get the \emph{initial} clusters. Then we encode these vertical clusters to \emph{hash} them into a new set of encoded configurations and apply CA again to get the \emph{primary} clusters. Finally, in Stage 3, we merge the clusters based on the desired number of clusters taken as user input. An initial version of this work is reported in Ref. \cite{abhishek2023cellular}. In \cite{manoranjan2023optimized}, the same algorithm is used over a new encoding technique with some preliminary characterization of the rules. In these works, the complexity of the clustering algorithm was $O(\mathcal{M}^3)$, that is, cubic, where $\mathcal{M} = |\mathcal{D}|$ is the number of elements in the dataset. Observe that, any clustering problem that takes a bijective mapping-based solution approach has the property that, its complexity can not be lower than $\Omega(\mathcal{M})$. So, any reversible CA-based solution can not have a complexity lower than that. In this work, which is an extension of both of Ref. \cite{abhishek2023cellular, manoranjan2023optimized}, we address the rule selection method vividly and find out the set of best rules considering the \emph{frequency based} encoding method \cite{dougherty1995supervised} used for converting data elements into binary. We also improve our clustering algorithm to drastically reduce the complexity of the technique making it quadratic time.

\section{Reversible Cellular Automata and Clustering}\label{sec:background}
This section gives an overview of how reversible CAs can be used for clustering. It also briefs about the state-of-the-art clustering algorithms and standard benchmark techniques used in the paper.
\par A \emph{cellular automaton}\cite{bhattacharjee2020survey} consists of a set of cells that are arranged as a regular network. Each cell of a CA is a finite automaton that uses a finite state set S. The CAs evolve in discrete time and space. During evolution, a cell of a CA changes its state depending on the present state of its neighbors. That is, to update its state, a cell uses a next state function, also known as a local rule, whose arguments are the present states of the cell’s neighbors. The collection of the states of all cells at a given time is called the configuration of the CA. During evolution, a CA, therefore, hops from one configuration to another. A CA is called a finite cellular automaton if the cellular space is finite. Finite CAs are really important if the automata are to be implemented. Two boundary conditions for finite CAs are generally used: the open boundary condition and the periodic boundary condition. Among the open boundary conditions, the most popular is the null boundary, where the missing neighbors of the terminal cells are always in state 0 (null). In this work, we are going to use finite one dimensional cellular automata under null boundary condition.

\subsection{Mapping Clustering with Cellular Automata}\label{sec:mappingCA}
Usually, the real-life datasets to be clustered are in the form of a set of data elements (rows) where each data element has several features (columns) with each feature's value as a real number. So, to apply CA, we first need to convert each of the data elements containing the feature (attribute) values into a binary string. For our work, we consider only the datasets with quantifiable feature properties such that they can be converted into binary strings, namely the datasets having only \emph{continuous} and \emph{categorical} attributes. However, the scheme will work on any other dataset provided the attribute values are intelligently encoded into binary numbers in the preprocessing stage.

This work considers {\em frequency based} encoding \cite{dougherty1995supervised} for the preprocessing. The essence of this encoding scheme is it converts the real numbers into binary maintaining a minimum \emph{hamming distance} between consecutive intervals. Here we consider that the range of values for the continuous attributes can be divided into a maximum of $4$ disjoint intervals, so two-bit encoding is required for such an attribute. Let $x$ be a feature value and ${\tt M}$ be the encoding function. Then
\[
{\tt M}(x) =
\begin{cases}
	00 & {x_1}\leq x\leq {x_s} \\
	01 & {x_{s+1}}\leq x\leq {x_w}\\
	11 & {x_{w+1}}\leq x\leq {x_q}\\
	10 & {x_{q+1}}\leq x\leq {x_t}\\
\end{cases}
\]
where [$x_1, x_s$], [$x_{s+1}, x_w$], [$x_{w+1}, x_q$]  and [$x_{q+1}, x_t$] are the maximum four disjoint intervals for the feature. Similarly, for a categorical attribute with $k$ distinct values, $k$ bits are needed to maintain minimum hamming distance where $i^{th}$ unique value will have one $1$ in the $i^{th}$ position with all other $k-1$ bits as $0$s. Once the features are encoded, the feature values of every attribute of an object can be concatenated to get a binary string per object. These binary strings can now be used in the CA as configurations of the reversible CA. We name these configurations as \emph{target configurations}.

As mentioned, we are considering $1$-dimensional $2$-state $2$-radius finite CAs for clustering. The reason for choosing radius (r) as $2$ is -- as per our encoding, a minimum of two bits are needed to represent a feature. In the CA, each cell updates its state based on a local rule, also called a \emph{rule} $\mathcal{R}: \{0,1\}^5 \rightarrow \{0,1\}$ which takes the neighborhood cells' values as arguments. Each of these combinations of arguments is also called a \emph{Rule Min Term} or an \emph{RMT}. It is often represented by the string generated by concatenating the arguments or its decimal equivalent. For example, the argument to $\mathcal{R}$ where all the neighbors have state $1$ is represented by RMT $11111$ or RMT $31$ which is the decimal equivalent of $11111$. Two RMTs can be grouped if, among the neighborhood combinations, all the neighbors except one have the same state. We can define it formally as follows:
\begin{definition}\label{def:equivalent}
	A set of RMTs is called \emph{k-equivalent} or $\mathcal{E}^k$, if the values of all neighbors of them are invariant except the $k^{th}$ neighbor, where $0\le k \le m-1$. Mathematically, $\mathcal{E}^k_i = \{ a_{1}a_{2}\cdots a_{k-1} {\tt x} a_{{k+1}}\cdots a_{m} \in \mathcal{S}^{m} ~|~ {\tt x} \in \mathcal{S}\}$. Here, $i$ is the decimal equivalent of ${a_{1}a_{2}\cdots a_{k-1} a_{{k+1}}\cdots a_{m}}$, $\mathcal{S}$ is set of states of the CA and $m$ is the number of neighbors. \cite{subrata_covid}.
\end{definition}

As we are taking $2$-state $2$-radius CA, that is, number of neighbors is $5$, we can get five such different $\mathcal{E}^k$, $0\le k \le 4$ where each $\mathcal{E}^k$ contains 16 sets $\mathcal{E}^k_i,~ 0\le i \le 15$. The groups are shown in Table~\ref{tab:equivalentRMT}.
\begin{table}[!ht]
	\centering
	\setlength{\tabcolsep}{4pt}
	\caption{The grouping of RMTs based on neighborhood equivalence}\label{tab:equivalentRMT}
	\begin{tabular}{|l|l||l|l||l|l||l|l||l|l|}
		\hline
		\multicolumn{2}{|c||}{$\mathcal{E}^4$} & \multicolumn{2}{c||}{$\mathcal{E}^3$} & \multicolumn{2}{c||}{$\mathcal{E}^2$} & \multicolumn{2}{c||}{$\mathcal{E}^1$} & \multicolumn{2}{c|}{$\mathcal{E}^0$} \\ \hline
		$\mathcal{E}_0^4$ & 0, 16 & $\mathcal{E}_0^3$ & 0,8 & $\mathcal{E}_0^2$ & 0,4 & $\mathcal{E}_0^1$ & 0,2 & $\mathcal{E}_0^0$ & 0,1 \\ \hline
		$\mathcal{E}_1^4$ & 1, 17 & $\mathcal{E}_1^3$ & 1,9 & $\mathcal{E}_1^2$ & 1,5 & $\mathcal{E}_1^1$ & 1,3 & $\mathcal{E}_1^0$ & 2,3 \\ \hline
		$\mathcal{E}_2^4$ & 2, 18 & $\mathcal{E}_2^3$ & 2,10 & $\mathcal{E}_2^2$ & 2,6 & $\mathcal{E}_2^1$ & 4,6 & $\mathcal{E}_2^0$ & 4,5 \\ \hline
		$\mathcal{E}_3^4$  & 3, 19 & $\mathcal{E}_3^3$ & 3,11 & $\mathcal{E}_3^2$ & 3,7 & $\mathcal{E}_3^1$ & 5,7 & $\mathcal{E}_3^0$ & 6,7 \\ \hline
		$\mathcal{E}_4^4$ & 4, 20 & $\mathcal{E}_4^3$ & 4,12 & $\mathcal{E}_4^2$ & 8,12 & $\mathcal{E}_4^1$ & 8,10 & $\mathcal{E}_4^0$ & 8,9 \\ \hline
		$\mathcal{E}_5^4$ & 5, 21 & $\mathcal{E}_5^3$ & 5,13 & 	$\mathcal{E}_5^2$ & 9,13 & 	$\mathcal{E}_5^1$ & 9,11 & 	$\mathcal{E}_5^0$ & 10,11 \\ \hline
		$\mathcal{E}_6^4$ & 6, 22 & $\mathcal{E}_6^3$ & 6,14 & 	$\mathcal{E}_6^2$ & 10,14 & $\mathcal{E}_6^1$ & 12,14 & $\mathcal{E}_6^0$ & 12, 13 \\ \hline
		$\mathcal{E}_7^4$ & 7, 23 & $\mathcal{E}_7^3$ & 7,15 & $\mathcal{E}_7^2$ & 11,15 & $\mathcal{E}_7^1$ & 13,15 & $\mathcal{E}_7^0$ & 14, 15 \\ \hline
		$\mathcal{E}_8^4$ & 8, 24 & $\mathcal{E}_8^3$ & 16,24 & $\mathcal{E}_8^2$ & 16,20 & $\mathcal{E}_8^1$ & 16,18 & $\mathcal{E}_8^0$ & 16, 17 \\ \hline
		$\mathcal{E}_9^4$ & 9, 25 & $\mathcal{E}_9^3$ & 17, 25 & $\mathcal{E}_9^2$ & 17,21 & $\mathcal{E}_9^1$ & 17,19 & $\mathcal{E}_9^0$ & 18, 19 \\ \hline
		$\mathcal{E}_{10}^4$ & 10, 26 & $\mathcal{E}_{10}^3$ & 18, 26 & $\mathcal{E}_{10}^2$ & 18,22 & $\mathcal{E}_{10}^1$ & 20,22 & $\mathcal{E}_{10}^0$ & 20, 21 \\ \hline
		$\mathcal{E}_{11}^4$ & 11, 27 & $\mathcal{E}_{11}^3$ & 19, 27 & $\mathcal{E}_{11}^2$ & 19,23 & $\mathcal{E}_{11}^1$ & 21,23 & $\mathcal{E}_{11}^0$ & 22, 23 \\ \hline
		$\mathcal{E}_{12}^4$ & 12, 28 & $\mathcal{E}_{12}^3$ & 20, 28 & $\mathcal{E}_{12}^2$ & 24,28 & $\mathcal{E}_{12}^1$ & 24,26 & $\mathcal{E}_{12}^0$ & 24,25 \\ \hline
		$\mathcal{E}_{13}^4$ & 13, 29 & $\mathcal{E}_{13}^3$ & 21, 29 & $\mathcal{E}_{13}^2$ & 25,29 & $\mathcal{E}_{13}^1$ & 25,27 & $\mathcal{E}_{13}^0$ & 26,27 \\ \hline
		$\mathcal{E}_{14}^4$ & 14, 20 & $\mathcal{E}_{14}^3$ & 22, 30 & $\mathcal{E}_{14}^2$ & 26,30 & $\mathcal{E}_{14}^1$ & 28,30 & $\mathcal{E}_{14}^0$ & 28,29 \\ \hline
		$\mathcal{E}_{15}^4$ & 15, 31 & $\mathcal{E}_{15}^3$ & 23, 31 & $\mathcal{E}_{15}^2$ & 27,31 & $\mathcal{E}_{15}^1$ & 29,31 & $\mathcal{E}_{15}^0$ & 30,31 \\ \hline
	\end{tabular}
\end{table}

A rule can be represented by the string generated by concatenating next state values of all RMTs starting from $31$ to $0$, that is, $\mathcal{R}(1,1,1,1,1)$ to $\mathcal{R}(0,0,0,0,0)$, or its decimal equivalent. For instance, rule $267422991$ represents the string $00001111111100001000110100001111$ which means, for this rule, $\mathcal{R}(0,0,0,0,0)=\mathcal{R}(0,0,0,0,1)=\mathcal{R}(0,0,0,1,0)=\mathcal{R}(0,0,0,1,1)=1$, $\mathcal{R}(0,0,1,0,0)=0$ and so on. A snapshot of the states of all cells at any time instant is called the configuration. The global transition function $G_n:\mathcal{C}_n\rightarrow \mathcal{C}_n$ works on the set of configurations $\mathcal{C}_n$: for any ${\tt x}, {\tt y} \in \mathcal{C}_n$, if ${\tt y}=({y_i})_{\forall i\in n}$ is the next configuration of ${\tt x} = (x_i)_{\forall i\in n}$, then ${\tt y}=G _n({\tt x}) = G_n (x_0x_1\cdots x_{n-1})= (\mathcal{R}(x_{i-2},x_{i-1}, x_i, x_{i+1}, x_{i+2}))_{0\le i\le n-1}$

\begin{definition}\label{Def:reachable}
	For any two configurations ${\tt x}, {\tt y} \in \mathcal{C}_n $, if ${\tt y} = G_n^k({\tt x})$ for some $k \in \mathbb{N}$, then ${\tt y}$ is called \emph{reachable} from ${\tt x}$; otherwise it is not reachable from ${{\tt x}}$ \cite{mukherjee2021reversible}
\end{definition}
\begin{definition}
	A CA is \emph{reversible} if each of its configurations is reachable from some other configurations.
\end{definition}

\begin{figure}[!h]
	\centering
	\includegraphics[scale=0.4]{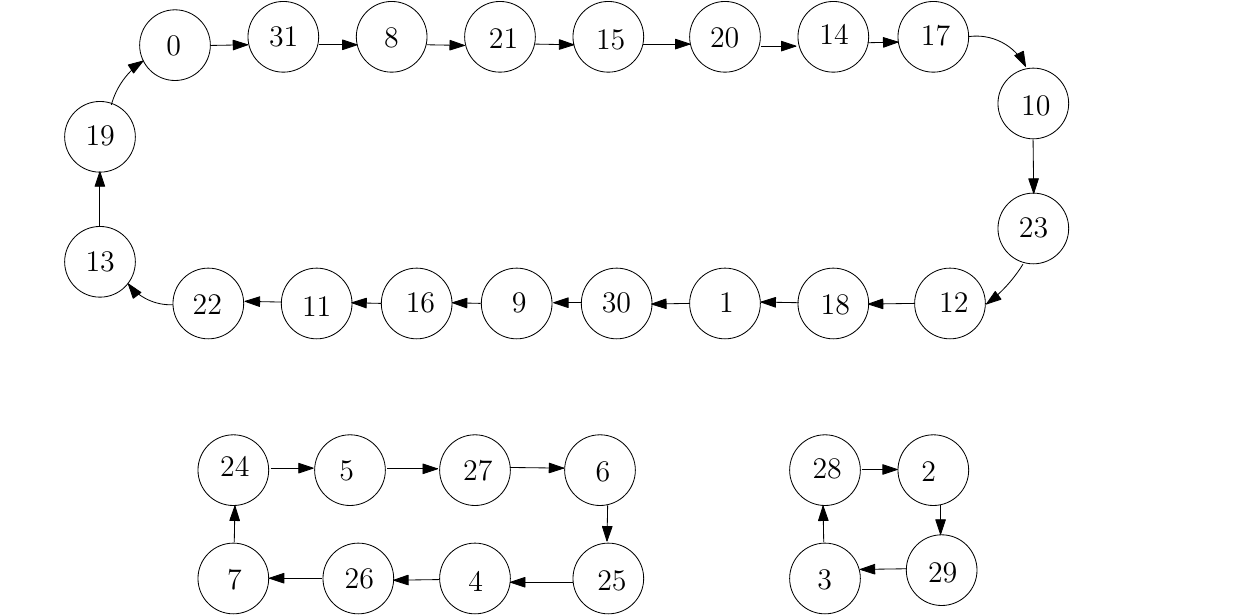}
	\caption{Evolution of a $5$-cell reversible CA $267422991$}
	\label{Fig:cycleEx}
\end{figure}

For example, Figure~\ref{Fig:cycleEx} represents the cycles of a $5$-cell reversible CA $267422991$. Here, the configurations are represented by their corresponding equivalent decimal numbers. We can see that, configuration $8 (01000)$ is reachable from the configuration $13 (01101)$ but not from $7 (00111)$. Now, let us consider a hypothetical dataset (shown in Table~\ref{tab:clustering_ex}) with one continuous and one categorical attribute. 
\begin{table}[h]
	\setlength{\tabcolsep}{1.4pt}
	\centering
	\caption{Frequency-based encoding of a hypothetical dataset}
	\label{tab:clustering_ex}
		\begin{tabular}{|c|c|c||c|c|c|}
			\hline
			\multirow{2}{*}{Objects}& \multicolumn{2}{c|}{Continuous Attribute} &\multicolumn{2}{c|}{Categorical Attribute}  & {Target Configuration} \\
			\hhline{~----~}
			&  No. of Petals & Encoding  & Color & Encoding & (with Decimal Equivalent)  \\
			\hline
			$Obj_1$ & 5 & 00 & White & 001 & 00001 (1) \\
			\hline
			$Obj_2$ & 10 & 01 & White & 001 & 01001 (9) \\
			\hline
			$Obj_3$ & 5 & 00 & Red & 010 & 00010 (2) \\
			\hline
			$Obj_4$ & 7 & 00 & Yellow & 100 & 00100 (4) \\
			\hline
			$Obj_5$ & 10 & 01 & Yellow &  100 & 01100 (12) \\
			\hline
			$Obj_6$ & 15 & 01 & Yellow & 100 & 01100 (12) \\
			\hline
			$Obj_7$ & 50 & 11 & White & 001 & 11001 (25) \\
			\hline
			$Obj_8$ & 55 & 11 & Red & 010 & 11010 (26) \\
			\hline
		\end{tabular}
\end{table} 

The categorical attribute has three distinct values, so we need three bits to encode it. Whereas, based on the values of the continuous attribute, we can create three disjoint sub-intervals $[5,7]$, $[10,15]$ and $[50,55]$ to be encoded by $00$, $01$, and $11$ respectively. In this way, each data element can be converted into a binary string which can be considered as the configurations of a $5$-cell CA. Observe that, the encoding function is surjective mapping more than one object into the same bits (see Object No. 5 (row 7) and 6 (row 8) of Table~\ref{tab:clustering_ex}): that is, the encoding itself provides a basic clustering of data elements. Now, let us cluster these objects using the CA $267422991$ of Figure~\ref{Fig:cycleEx}. As per the cycles of this CA, the configurations $1$, $9$, and $12$ are reachable from each other and form one cluster, whereas, $4, 25$ \& $26$ form another cluster and $2$ belong to a third cluster. So, by CA $267422991$, the objects of this hypothetical dataset are clustered as $(Obj_1, Obj_2, Obj_5, Obj_6)$, $(Obj_3)$, $(Obj_4, Obj_7, Obj_8)$.

\subsection{Clustering Algorithms and Performance Metrics}
Let us first briefly discuss some established algorithms that are to be used as a reference point of comparison in this paper. These algorithms are well-documented and work for both larger and smaller datasets. The most classical of these algorithms is $K$-means \cite{hartigan1979algorithm, likas2003global}.  Introduced in 1973 and inspiring researchers to come up with improvements since then, this seminal algorithm figures out $K$ number of center points in a data set and then groups each data point into the nearest centroid. The next popular algorithm is DBSCAN \cite{ester1996density}. It works on the principle of density-based spatial clustering of applications with noise where the neighborhood within a given radius has to have at least a certain number of points for each point of a cluster. A contemporary algorithm to DBSCAN is BIRCH \cite{zhang1997birch} which clusters huge datasets by first producing a compact summary of a large dataset while maintaining as much information available after which the clustering takes place. A relatively new algorithm is Meanshift \cite{comaniciu2002mean}. This algorithm uses a pattern recognition procedure, mean shift to assign data points to clusters in an iterative manner while shifting the data points toward the modes of density. The most effective one is agglomerative hierarchical clustering, here abbreviated as \emph{hierarchical} \cite{JSSv025i04}, which iteratively clusters the data points giving the best set of clusters.

\begin{table}[!hbtp]
	\centering

	\caption{Performance Metrics}\label{tab2}
	
		\begin{tabular}{|l|p{8cm}|}
			\hline
			\bfseries Metrics &  \bfseries Formula \\
			\hline
			
			\bfseries Silhouette Score\cite{sil} & {$\mathbf{{Score = (b-a)/\max(a,b)}}$ } \\
			&\\
			\bfseries Davis-Bouldin\cite{ester1996density} &  {$\mathbf{DB = \frac{1}{n_c}\sum\limits_{i=1}^{n_c}R_i}$, where $\mathbf{R_i = \max\limits_{j=1, \cdots, n_c, i\ne j}{R_{ij}}}$, $\mathbf{i=1, \cdots, n_c}$} \\
			&\\
			\bfseries Calinski-Harabasz\cite{calinski1974dendrite} & {$\mathbf{CH = \left[\frac{\sum\limits_{k=1}^{K}n_k {\lVert{c_k-c}\rVert^2}}{K-1}\right]}$/ $\mathbf{\left[\frac{\sum\limits_{k=1}^{K}\sum\limits_{i=1}^{n_k}{\lVert d_i-c_k\rVert}^2}{N-K}\right]}$}\\
			\hline
	\end{tabular}

\end{table}
Table~\ref{tab2} lists the standard benchmark validation indices that we are using for checking the performance of the clustering algorithms. Here, the Silhouette score is a real number within (-1, 1) where the higher the score, the better the clusters. The same is also true for the Calinski-Harabasz indexing but without any fixed range. However, in the case of the Davies-Bouldin index, a lower score indicates better clustering with the ideal score as zero.

\section{Selection of Proper Rules}\label{sec:rule_selection}
Section~\ref{sec:mappingCA} gives a basic idea of clustering using a reversible CA. However, any clustering is \emph{good} if the \emph{distance} between the elements in any pair of clusters is very large in comparison to that for the elements within a cluster. So, while selecting a reversible CA for the clustering, we need to satisfy that, for the chosen CA, the configurations that belong to the same cycle maintain this minimum distance property while the distance between a pair of configurations from two different cycles is very large. This section describes our method of looking for the CAs which hold this property and gives the final list of candidate CAs suitable for clustering any dataset under the frequency-based encoding method.

\subsection{Filtering based on Information Propagation}
For $2$-state $5$-neighborhood CAs under null boundary condition, there are a total of $226$ rules which are reversible for some $n\in \mathbb{N}$  \cite{mukherjee2021reversible}. Among these CAs, our first criterion is to filter out the CAs having a strict \emph{locality} property. This indicates that, for these CAs, at the time of evolution, there is not much change in the states of the cells while hopping from one configuration to the next. This can happen when there is less amount of \emph{information propagation} from one configuration to the next. Also, the rule needs to have a high rate of \emph{self-replication}, that is, for the rule, there are many RMTs ($x_{i-2} x_{i-1} x_i x_{i+1} x_{i+2}$), for which the next state is the state of the cell itself, that is, $\mathcal{R}(x_{i-2},x_{i-1},x_i,x_{i+1}, x_{i+2}) = x_i$. Here, $x_i$ denotes the state of the $i^{th}$ cell. When the configuration contains a cell with such a neighborhood combination, the rule keeps the state of the cell unchanged. Because of these, the configurations that are \emph{similar} belong to the same cycle. 

Now, this information propagation in the CA is calculated with respect to every neighbor. For that, we take each pair of arguments to the rule that differs by only one neighborhood position and see if the corresponding next states are different. According to Definition~\ref{def:equivalent}, with respect to the $k^{th}$ neighbor, $\mathcal{E}^{k}_i$, for each $i$, is such a set. If the next state values of the RMTs of $\mathcal{E}^{k}_i$ are different, then that neighborhood combination plays no role in updating the state of the cell under consideration. For example, if $\mathcal{R}(x_{i-2},0,x_i,x_{i+1}, x_{i+2}) = \mathcal{R}(x_{i-2},1, x_i,x_{i+1}, x_{i+2})$ for all values of $x_{i-2},x_i,x_{i+1}, x_{i+2}$, that means, the next state of $i^{th}$ cell is independent of any changes in its left neighbor. 	
The cumulative sum of all such pairs per neighbor can be used to derive the rate of information propagation for that neighbor. For instance, in the above example, the rate of information propagation from this neighbor is zero. The detailed derivation of the formula used for calculating it can be found in Ref.~\cite{Supreeti_2018_chaos, subrata_covid}. In short, for each neighbor $k$, $0\le k \le 4$, information propagation to the $k^{th}$ neighbor due to change in the current cell is 
\[\Lambda^k_i  = \frac{1}{2^4}\sum_{i=0}^{2^4-1}\lambda^k_i = \frac{1}{16}\sum_{i=0}^{15}\lambda^k_i\]
where 
\[\lambda^k_i = \frac{1}{2}\sum_{r,s \in \mathcal{E}^{k}_i, r \ne s}\delta^k_i(r,s)\] and \[\delta^k_i(r,s)  =
\begin{cases}
	1 & \text{if $R[r] \ne R[s]$ where $r,s \in \mathcal{E}^{k}_i,~ r \ne s$} \\
	0 & \text{otherwise}
\end{cases}\]

The information propagation rate considering the cell itself as its neighbor, that is, for $k=2$, is the self-replication rate which is the information propagation for the cell itself.
The maximum information propagation rate for any neighbor is 100\% and the minimum is 0\%. To choose the rules having strict locality property means, the rules need to have high rate of self-replication and a low rate of information propagation. Hence, our first filtering criterion is:\\
\textbf{Criterion 1:} \emph{Select the CA rules such that the rate of information propagation for each neighbor is $\le 75\%$ but not all of them are equal to $0$ or $75\%$ and the rate of self-replication is $\ge 75\%$.}


\begin{center}
	\begin{small}\setlength\tabcolsep{1.5pt}
		\begin{longtable}{|c|c||c|c|c|c|c|}
			\caption{$162$ Potential Candidate CA rules based on Information Propagation}\label{tab:revRules}\\
			\hline
			\multicolumn{2}{|c||}{\textbf{Rule}} & \multicolumn{5}{c|}{\textbf{Information Propagation Rate}}\\
			\hline
			Binary & Decimal & $i-2$ & $i-1$ & $i$ & $i+1$ & $i+2$ \\
			\hline
			\endfirsthead
			\multicolumn{7}{c}%
			{\tablename\ \thetable\ -- \textit{Continued from previous page}} \\
			\hline
			\multicolumn{2}{|c||}{\textbf{Rule}} & \multicolumn{5}{c|}{\textbf{Information Propagation Rate}}\\
			\hline
			Binary & Decimal & $i-2$ & $i-1$ & $i$ & $i+1$ & $i+2$ \\
			\hline
			\endhead
			\hline \multicolumn{7}{r}{\textit{Continued on next page}} \\
			\endfoot
			\hline
			\endlastfoot
			00001111000011110000111101001011 & 252645195 & 12.5 & 12.5 & 100.0 & 12.5 & 12.5 \\
			00001111000011110000111111011000 & 252645336 & 37.5 & 37.5 & 87.5 & 12.5 & 12.5 \\
			00001111000011110000111111110000 & 252645360 & 50.0 & 50.0 & 100.0 & 0.0 & 0.0 \\
			00001111000011110001111000001111 & 252648975 & 12.5 & 12.5 & 100.0 & 12.5 & 12.5 \\
			00001111000011110010110100001111 & 252652815 & 12.5 & 12.5 & 100.0 & 12.5 & 12.5 \\
			00001111000011110011110000001111 & 252656655 & 25.0 & 25.0 & 100.0 & 25.0 & 0.0 \\
			00001111000011110011110011110000 & 252656880 & 75.0 & 25.0 & 100.0 & 25.0 & 0.0 \\
			00001111000011111100001100001111 & 252691215 & 25.0 & 25.0 & 100.0 & 25.0 & 0.0 \\
			00001111000011111100001111110000 & 252691440 & 75.0 & 25.0 & 100.0 & 25.0 & 0.0 \\
			00001111000011111101001000001111 & 252695055 & 37.5 & 37.5 & 100.0 & 12.5 & 12.5 \\
			00001111000011111110000100001111 & 252698895 & 37.5 & 37.5 & 100.0 & 12.5 & 12.5 \\
			00001111000011111111000000001111 & 252702735 & 50.0 & 50.0 & 100.0 & 0.0 & 0.0 \\
			00001111000111101101001000011110 & 253678110 & 37.5 & 37.5 & 100.0 & 37.5 & 37.5 \\
			00001111001011010000111100101101 & 254611245 & 0.0 & 25.0 & 100.0 & 25.0 & 25.0 \\
			00001111001011010010110100101101 & 254618925 & 12.5 & 12.5 & 100.0 & 37.5 & 37.5 \\
			00001111001111000000111100001111 & 255594255 & 25.0 & 25.0 & 100.0 & 25.0 & 0.0 \\
			00001111001111001111000011110000 & 255652080 & 75.0 & 25.0 & 100.0 & 25.0 & 0.0 \\
			00001111010010110000111100001111 & 256577295 & 12.5 & 12.5 & 100.0 & 12.5 & 12.5 \\
			00001111010010110000111101001011 & 256577355 & 0.0 & 25.0 & 100.0 & 25.0 & 25.0 \\
			00001111011110000000111100001111 & 259526415 & 37.5 & 37.5 & 100.0 & 12.5 & 12.5 \\
			00001111100001110000111100001111 & 260509455 & 12.5 & 12.5 & 100.0 & 12.5 & 12.5 \\
			00001111100011011111000000001111 & 260960271 & 62.5 & 62.5 & 87.5 & 12.5 & 12.5 \\
			00001111101101000000111100001111 & 263458575 & 37.5 & 37.5 & 100.0 & 12.5 & 12.5 \\
			00001111101101000000111101001011 & 263458635 & 50.0 & 50.0 & 100.0 & 25.0 & 25.0 \\
			00001111110000110000111100001111 & 264441615 & 25.0 & 25.0 & 100.0 & 25.0 & 0.0 \\
			00001111110000111111000011110000 & 264499440 & 75.0 & 25.0 & 100.0 & 25.0 & 0.0 \\
			00001111110100101111000011010010 & 265482450 & 50.0 & 50.0 & 100.0 & 25.0 & 25.0 \\
			00001111111100000000111100001111 & 267390735 & 50.0 & 50.0 & 100.0 & 0.0 & 0.0 \\
			00001111111100000000111101001011 & 267390795 & 37.5 & 62.5 & 100.0 & 12.5 & 12.5 \\
			00001111111100001000110100001111 & 267422991 & 62.5 & 62.5 & 87.5 & 12.5 & 12.5 \\
			00001111111100001111000011110000 & 267448560 & 50.0 & 50.0 & 100.0 & 0.0 & 0.0 \\
			00011110000111100001111000011110 & 505290270 & 0.0 & 0.0 & 100.0 & 50.0 & 50.0 \\
			00011110000111101101001000011110 & 505336350 & 25.0 & 25.0 & 100.0 & 50.0 & 50.0 \\
			00011110010110100001111001011010 & 509222490 & 0.0 & 25.0 & 100.0 & 25.0 & 75.0 \\
			00011110110100101110000111010010 & 517136850 & 50.0 & 50.0 & 100.0 & 50.0 & 50.0 \\
			00011110110100101111000011010010 & 517140690 & 37.5 & 37.5 & 100.0 & 37.5 & 37.5 \\
			00011111000011100001111000011110 & 521018910 & 12.5 & 12.5 & 87.5 & 37.5 & 37.5 \\
			00100001110111101101001011010010 & 568251090 & 50.0 & 50.0 & 75.0 & 50.0 & 50.0 \\
			00100001111111000010110111110000 & 570174960 & 25.0 & 75.0 & 75.0 & 37.5 & 25.0 \\
			00100001111111001101001011110000 & 570217200 & 50.0 & 50.0 & 75.0 & 37.5 & 25.0 \\
			00101001011011010110100101101001 & 695036265 & 12.5 & 12.5 & 87.5 & 87.5 & 87.5 \\
			00101101000011110000111100001111 & 755961615 & 12.5 & 12.5 & 100.0 & 12.5 & 12.5 \\
			00101101000011110010110100001111 & 755969295 & 0.0 & 25.0 & 100.0 & 25.0 & 25.0 \\
			00101101000011111101001000001111 & 756011535 & 50.0 & 50.0 & 100.0 & 25.0 & 25.0 \\
			00101101000011111111000000001111 & 756019215 & 37.5 & 62.5 & 100.0 & 12.5 & 12.5 \\
			00101101000111101101001000011110 & 756994590 & 50.0 & 50.0 & 100.0 & 50.0 & 50.0 \\
			00101101001011010000111100101101 & 757927725 & 12.5 & 12.5 & 100.0 & 37.5 & 37.5 \\
			00101101001011010001111000101101 & 757931565 & 25.0 & 25.0 & 100.0 & 50.0 & 50.0 \\
			00101101001011010010110100101101 & 757935405 & 0.0 & 0.0 & 100.0 & 50.0 & 50.0 \\
			00101101001011010011110000101101 & 757939245 & 12.5 & 12.5 & 100.0 & 62.5 & 37.5 \\
			00101101100011011100001110001101 & 764265357 & 37.5 & 37.5 & 75.0 & 62.5 & 37.5 \\
			00101101100011011101001010001101 & 764269197 & 50.0 & 50.0 & 75.0 & 50.0 & 50.0 \\
			00101101100011011110000110001101 & 764273037 & 25.0 & 37.5 & 75.0 & 50.0 & 50.0 \\
			00101101100011011111000010001101 & 764276877 & 37.5 & 50.0 & 75.0 & 37.5 & 37.5 \\
			00110110111100000011011000111100 & 921712188 & 25.0 & 37.5 & 75.0 & 75.0 & 25.0 \\
			00111100100111001111000010011100 & 1016918172 & 25.0 & 37.5 & 75.0 & 75.0 & 25.0 \\
			01001011000011110100101100001111 & 1259293455 & 0.0 & 25.0 & 100.0 & 25.0 & 25.0 \\
			01001011000011110100101101001011 & 1259293515 & 12.5 & 12.5 & 100.0 & 37.5 & 37.5 \\
			01001011000011110100101101111000 & 1259293560 & 37.5 & 37.5 & 100.0 & 37.5 & 37.5 \\
			01001011000011110100101111110000 & 1259293680 & 50.0 & 50.0 & 100.0 & 25.0 & 25.0 \\
			01001011010010110100101100001111 & 1263225615 & 12.5 & 12.5 & 100.0 & 37.5 & 37.5 \\
			01001011010010110100101101001011 & 1263225675 & 0.0 & 0.0 & 100.0 & 50.0 & 50.0 \\
			01001011100001110100101101001011 & 1267157835 & 25.0 & 25.0 & 100.0 & 50.0 & 50.0 \\
			01001011100001110100101101111000 & 1267157880 & 50.0 & 50.0 & 100.0 & 50.0 & 50.0 \\
			01001011110000110100101101001011 & 1271089995 & 12.5 & 12.5 & 100.0 & 62.5 & 37.5 \\
			01011010000111100101101000011110 & 1511938590 & 0.0 & 25.0 & 100.0 & 25.0 & 75.0 \\
			01011010011110000101101001111000 & 1517836920 & 0.0 & 25.0 & 100.0 & 25.0 & 75.0 \\
			01101011010010010100101101001011 & 1799965515 & 12.5 & 12.5 & 87.5 & 62.5 & 62.5 \\
			01101101001010010010110100101101 & 1831415085 & 12.5 & 12.5 & 87.5 & 62.5 & 62.5 \\
			01110010100001110111001001111000 & 1921479288 & 50.0 & 50.0 & 75.0 & 50.0 & 50.0 \\
			01110010101101000111001001111000 & 1924428408 & 25.0 & 37.5 & 75.0 & 50.0 & 50.0 \\
			01111000010010110111100001111000 & 2018211960 & 25.0 & 25.0 & 100.0 & 50.0 & 50.0 \\
			01111000010010110111100010110100 & 2018212020 & 50.0 & 50.0 & 100.0 & 50.0 & 50.0 \\
			01111000010010110111100011110000 & 2018212080 & 37.5 & 37.5 & 100.0 & 37.5 & 37.5 \\
			01111000010110100111100001011010 & 2019194970 & 0.0 & 25.0 & 100.0 & 25.0 & 75.0 \\
			01111000011110000111100001111000 & 2021161080 & 0.0 & 0.0 & 100.0 & 50.0 & 50.0 \\
			01111011100001000100101100001111 & 2072267535 & 37.5 & 62.5 & 75.0 & 37.5 & 37.5 \\
			01111011100001000100101101001011 & 2072267595 & 50.0 & 50.0 & 75.0 & 50.0 & 50.0 \\
			01111011110000000100101100001111 & 2076199695 & 50.0 & 50.0 & 75.0 & 37.5 & 25.0 \\
			01111011110000000100101101001011 & 2076199755 & 37.5 & 37.5 & 75.0 & 50.0 & 37.5 \\
			01111011110000000100101111110000 & 2076199920 & 25.0 & 75.0 & 75.0 & 37.5 & 25.0 \\
			10000100001111111011010000001111 & 2218767375 & 25.0 & 75.0 & 75.0 & 37.5 & 25.0 \\
			10000100001111111011010010110100 & 2218767540 & 37.5 & 37.5 & 75.0 & 50.0 & 37.5 \\
			10000100001111111011010011110000 & 2218767600 & 50.0 & 50.0 & 75.0 & 37.5 & 25.0 \\
			10000100011110111011010010110100 & 2222699700 & 50.0 & 50.0 & 75.0 & 50.0 & 50.0 \\
			10000100011110111011010011110000 & 2222699760 & 37.5 & 62.5 & 75.0 & 37.5 & 37.5 \\
			10000111100001111000011110000111 & 2273806215 & 0.0 & 0.0 & 100.0 & 50.0 & 50.0 \\
			10000111101001011000011110100101 & 2275772325 & 0.0 & 25.0 & 100.0 & 25.0 & 75.0 \\
			10000111101101001000011100001111 & 2276755215 & 37.5 & 37.5 & 100.0 & 37.5 & 37.5 \\
			10000111101101001000011101001011 & 2276755275 & 50.0 & 50.0 & 100.0 & 50.0 & 50.0 \\
			10000111101101001000011110000111 & 2276755335 & 25.0 & 25.0 & 100.0 & 50.0 & 50.0 \\
			10001101010010111000110110000111 & 2370538887 & 25.0 & 37.5 & 75.0 & 50.0 & 50.0 \\
			10001101011110001000110110000111 & 2373488007 & 50.0 & 50.0 & 75.0 & 50.0 & 50.0 \\
			10010010110101101101001011010010 & 2463552210 & 12.5 & 12.5 & 87.5 & 62.5 & 62.5 \\
			10010100101101101011010010110100 & 2495001780 & 12.5 & 12.5 & 87.5 & 62.5 & 62.5 \\
			10100101100001111010010110000111 & 2777130375 & 0.0 & 25.0 & 100.0 & 25.0 & 75.0 \\
			10100101111000011010010111100001 & 2783028705 & 0.0 & 25.0 & 100.0 & 25.0 & 75.0 \\
			10110100001111001011010010110100 & 3023877300 & 12.5 & 12.5 & 100.0 & 62.5 & 37.5 \\
			10110100011110001011010010000111 & 3027809415 & 50.0 & 50.0 & 100.0 & 50.0 & 50.0 \\
			10110100011110001011010010110100 & 3027809460 & 25.0 & 25.0 & 100.0 & 50.0 & 50.0 \\
			10110100101101001011010010110100 & 3031741620 & 0.0 & 0.0 & 100.0 & 50.0 & 50.0 \\
			10110100101101001011010011110000 & 3031741680 & 12.5 & 12.5 & 100.0 & 37.5 & 37.5 \\
			10110100111100001011010000001111 & 3035673615 & 50.0 & 50.0 & 100.0 & 25.0 & 25.0 \\
			10110100111100001011010010000111 & 3035673735 & 37.5 & 37.5 & 100.0 & 37.5 & 37.5 \\
			10110100111100001011010010110100 & 3035673780 & 12.5 & 12.5 & 100.0 & 37.5 & 37.5 \\
			10110100111100001011010011110000 & 3035673840 & 0.0 & 25.0 & 100.0 & 25.0 & 25.0 \\
			10110110100101001001011010010110 & 3063191190 & 12.5 & 12.5 & 87.5 & 87.5 & 87.5 \\
			11000011011000110000111101100011 & 3278049123 & 25.0 & 37.5 & 75.0 & 75.0 & 25.0 \\
			11001001000011111100100111000011 & 3373255107 & 25.0 & 37.5 & 75.0 & 75.0 & 25.0 \\
			11010010011100100000111101110010 & 3530690418 & 37.5 & 50.0 & 75.0 & 37.5 & 37.5 \\
			11010010011100100001111001110010 & 3530694258 & 25.0 & 37.5 & 75.0 & 50.0 & 50.0 \\
			11010010011100100010110101110010 & 3530698098 & 50.0 & 50.0 & 75.0 & 50.0 & 50.0 \\
			11010010011100100011110001110010 & 3530701938 & 37.5 & 37.5 & 75.0 & 62.5 & 37.5 \\
			11010010110100101100001111010010 & 3537028050 & 12.5 & 12.5 & 100.0 & 62.5 & 37.5 \\
			11010010110100101101001011010010 & 3537031890 & 0.0 & 0.0 & 100.0 & 50.0 & 50.0 \\
			11010010110100101110000111010010 & 3537035730 & 25.0 & 25.0 & 100.0 & 50.0 & 50.0 \\
			11010010110100101111000011010010 & 3537039570 & 12.5 & 12.5 & 100.0 & 37.5 & 37.5 \\
			11010010111000010010110111100001 & 3537972705 & 50.0 & 50.0 & 100.0 & 50.0 & 50.0 \\
			11010010111100000000111111110000 & 3538948080 & 37.5 & 62.5 & 100.0 & 12.5 & 12.5 \\
			11010010111100000010110111110000 & 3538955760 & 50.0 & 50.0 & 100.0 & 25.0 & 25.0 \\
			11010010111100001101001011110000 & 3538998000 & 0.0 & 25.0 & 100.0 & 25.0 & 25.0 \\
			11010010111100001111000011110000 & 3539005680 & 12.5 & 12.5 & 100.0 & 12.5 & 12.5 \\
			11011110000000110010110100001111 & 3724750095 & 50.0 & 50.0 & 75.0 & 37.5 & 25.0 \\
			11011110000000111101001000001111 & 3724792335 & 25.0 & 75.0 & 75.0 & 37.5 & 25.0 \\
			11011110001000010010110100101101 & 3726716205 & 50.0 & 50.0 & 75.0 & 50.0 & 50.0 \\
			11100000111100011110000111100001 & 3773948385 & 12.5 & 12.5 & 87.5 & 37.5 & 37.5 \\
			11100001001011010000111100101101 & 3777826605 & 37.5 & 37.5 & 100.0 & 37.5 & 37.5 \\
			11100001001011010001111000101101 & 3777830445 & 50.0 & 50.0 & 100.0 & 50.0 & 50.0 \\
			11100001101001011110000110100101 & 3785744805 & 0.0 & 25.0 & 100.0 & 25.0 & 75.0 \\
			11100001111000010010110111100001 & 3789630945 & 25.0 & 25.0 & 100.0 & 50.0 & 50.0 \\
			11100001111000011110000111100001 & 3789677025 & 0.0 & 0.0 & 100.0 & 50.0 & 50.0 \\
			11110000000011110000111100001111 & 4027518735 & 50.0 & 50.0 & 100.0 & 0.0 & 0.0 \\
			11110000000011110111001011110000 & 4027544304 & 62.5 & 62.5 & 87.5 & 12.5 & 12.5 \\
			11110000000011111111000010110100 & 4027576500 & 37.5 & 62.5 & 100.0 & 12.5 & 12.5 \\
			11110000000011111111000011110000 & 4027576560 & 50.0 & 50.0 & 100.0 & 0.0 & 0.0 \\
			11110000001011010000111100101101 & 4029484845 & 50.0 & 50.0 & 100.0 & 25.0 & 25.0 \\
			11110000001111000000111100001111 & 4030467855 & 75.0 & 25.0 & 100.0 & 25.0 & 0.0 \\
			11110000001111001111000011110000 & 4030525680 & 25.0 & 25.0 & 100.0 & 25.0 & 0.0 \\
			11110000010010111111000010110100 & 4031508660 & 50.0 & 50.0 & 100.0 & 25.0 & 25.0 \\
			11110000010010111111000011110000 & 4031508720 & 37.5 & 37.5 & 100.0 & 12.5 & 12.5 \\
			11110000011100100000111111110000 & 4034007024 & 62.5 & 62.5 & 87.5 & 12.5 & 12.5 \\
			11110000011110001111000011110000 & 4034457840 & 12.5 & 12.5 & 100.0 & 12.5 & 12.5 \\
			11110000100001111111000011110000 & 4035440880 & 37.5 & 37.5 & 100.0 & 12.5 & 12.5 \\
			11110000101101001111000010110100 & 4038389940 & 0.0 & 25.0 & 100.0 & 25.0 & 25.0 \\
			11110000101101001111000011110000 & 4038390000 & 12.5 & 12.5 & 100.0 & 12.5 & 12.5 \\
			11110000110000110000111100001111 & 4039315215 & 75.0 & 25.0 & 100.0 & 25.0 & 0.0 \\
			11110000110000111111000011110000 & 4039373040 & 25.0 & 25.0 & 100.0 & 25.0 & 0.0 \\
			11110000110100101101001011010010 & 4040348370 & 12.5 & 12.5 & 100.0 & 37.5 & 37.5 \\
			11110000110100101111000011010010 & 4040356050 & 0.0 & 25.0 & 100.0 & 25.0 & 25.0 \\
			11110000111000010010110111100001 & 4041289185 & 37.5 & 37.5 & 100.0 & 37.5 & 37.5 \\
			11110000111100000000111111110000 & 4042264560 & 50.0 & 50.0 & 100.0 & 0.0 & 0.0 \\
			11110000111100000001111011110000 & 4042268400 & 37.5 & 37.5 & 100.0 & 12.5 & 12.5 \\
			11110000111100000010110111110000 & 4042272240 & 37.5 & 37.5 & 100.0 & 12.5 & 12.5 \\
			11110000111100000011110000001111 & 4042275855 & 75.0 & 25.0 & 100.0 & 25.0 & 0.0 \\
			11110000111100000011110011110000 & 4042276080 & 25.0 & 25.0 & 100.0 & 25.0 & 0.0 \\
			11110000111100001100001100001111 & 4042310415 & 75.0 & 25.0 & 100.0 & 25.0 & 0.0 \\
			11110000111100001100001111110000 & 4042310640 & 25.0 & 25.0 & 100.0 & 25.0 & 0.0 \\
			11110000111100001101001011110000 & 4042314480 & 12.5 & 12.5 & 100.0 & 12.5 & 12.5 \\
			11110000111100001110000111110000 & 4042318320 & 12.5 & 12.5 & 100.0 & 12.5 & 12.5 \\
			11110000111100001111000000001111 & 4042321935 & 50.0 & 50.0 & 100.0 & 0.0 & 0.0 \\
			11110000111100001111000000100111 & 4042321959 & 37.5 & 37.5 & 87.5 & 12.5 & 12.5 \\
			11110000111100001111000010110100 & 4042322100 & 12.5 & 12.5 & 100.0 & 12.5 & 12.5 \\
			\hline
		\end{longtable}
	\end{small}
\end{center}

Table~\ref{tab:revRules} lists the $162$ possible candidate rules out of the total 226 rules, which satisfy our criteria. Here, the first column enlists the rule in decimal value and the next five columns note down the rate of information propagation (in percentage) for that rule concerning each of the five neighbors. All of these rules are reversible for all $n \in \mathbb{N}$. But $162$ is still a huge number. So, we need to apply stricter criteria to these rules by analyzing the inherent cycle structure of the CAs to further reduce the candidate rule set size.

\subsection{Filtering based on Cycle Structure}
A good clustering implies similar data points are placed together. In the case of CA, as cycles are instrumental in deciding which data points will be in the same cluster, we need to ensure that, in the chosen CA, the configurations inside each cycle maintain a minimum distance. As we have chosen frequency-based encoding, the objects have been encoded maintaining minimum hamming distance between consecutive ranges. So, as per this encoding, the target configurations (encoded objects) which have minimum hamming distance between them, need to be placed in the same cluster to make the clustering effective. This inevitably indicates, that in our CA, the hamming distance between the configurations of each cycle needs to be as minimum as possible. 

To maintain this, we calculate the \emph{average hamming distance ($\mathcal{H}_{avg}$)} between the configurations of each cycle. As the configurations are in binary, this average hamming distance is just bit-wise XOR of the configurations. So, for an $n$-cell CA with number of cycles $\mathcal{K}$, where cycle number $i$ ($cycle_i$) contains $k_i$ elements $c^i_1, c^i_2, \cdots, c^i_{k_i}$, the average hamming distance ($\mathcal{H}^i_{avg}$) is calculated as:
\[\mathcal{H}^i_{avg} = c^i_1 \oplus c^i_2 \oplus \cdots \oplus c^i_{k_i}\]

We need our rules to have this $\mathcal{H}^i_{avg}$ to be as small as possible for all $i\in \mathcal{K}$. The minimum value is $0$. However, if we restrict the rules to have all $\mathcal{H}^i_{avg}=0$ for every $i\in \mathcal{K}$, then we are left with no rule in our selected $162$ rules. If we take the CAs to have all $\mathcal{H}^i_{avg}\le 9$ for every $i\in \mathcal{K}$, then for $n=13$, we have only one rule $4042321935$ where out of the total $56$ cycles, average hamming distance is $0$ for each of the $48$ cycles and $1$ for each of the remaining $8$ cycles. So, restricting all $\mathcal{H}^i_{avg}\le 9$ is not practical. We need to have a good number of cycles with this low average hamming distance such that there are enough rules in the list. This leads us to our second filtering criterion: \\
\textbf{Criterion 2:} \emph{For any $n$, choose those CA rules out of Table~\ref{tab:revRules} for which there exists at least ${l}_1$ percent of cycles in the CA, such that in each of those cycles ($i$), the average hamming distance $\mathcal{H}^i_{avg}\le 9$.}

This $l_1$ can be set according to the user's requirement. For example, considering $l_1=40\%$ for $n=13$, we get a list of $45$ CA rules. This list is shown in Table~\ref{tab:criteria23}. Similarly, for other $n$, we can derive the list. For this work, the value of $l_1$ chosen for different $n$ along with the number of rules satisfying that criterion is shown in Table~\ref{tab:criteria23_rulesize} Now, although these rules have a good number of cycles maintaining minimum hamming distance, many of them have a very large number of cycles. To achieve the desired number of clusters in less time, we must have some CAs with a limited number of cycles. So, to include such CAs, we relax our constraint on the hamming distance a little -- we allow CAs with a limited number of cycles such that at least 50\% of the cycles have an average hamming distance of only two digits. Therefore, our third and final selection criteria is:\\ 
\textbf{Criterion 3:} \emph{For any $n$, choose those CA rules out of Table~\ref{tab:revRules} for which number of cycles in the CA is $\le l_2$ and at least 50\% of these cycles have the average hamming distance $\mathcal{H}_{avg}\le 99$.}

\begin{table}[hbtp]
	\centering
	\setlength{\tabcolsep}{1.8pt}
	\caption{Selected Candidates for $n=13$ from $162$ Rules of Table~\ref{tab:revRules}}\label{tab:criteria23}
	\resizebox*{0.8\textwidth}{!}{
	\begin{tabular}{|c|c|c|c|c||c|}
		\hline
		\multicolumn{5}{|c||}{\textbf{Based on only Criteria 2}} & {\textbf{Based on Criteria 3}}\\
		\hline
		1259293455 & 252648975 & 255652080 & 3031741620 & 505290270 & 4042310415 \\ \hline
		1263225615 & 252656655 & 256577355 & 3789677025 & 517140690 & \textbf{252691440} \\ \hline
		1263225675 & 252691215 & 259526415 & 4027544304 & 521018910 & 267422991 \\ \hline
		1921479288 & \textbf{252691440} & 260509455 & 4027576560 & 755961615 & 4039315215 \\ \hline
		2018211960 & 252698895 & 264441615 & 4034007024 & 755969295 & ~ \\ \hhline{-----~}
		2018212080 & 252702735 & 264499440 & 4035440880 & 757935405 & ~ \\ \hhline{-----~}
		2273806215 & 254611245 & 265482450 & 4041289185 & 4027518735 & ~ \\ \hhline{-----~}
		252645195 & 254618925 & 267390735 & 4042264560 & 4031508720 & ~ \\ \hhline{-----~}
		252645360 & 255594255 & 2783028705 & 4042272240 & 4042321935 &\\ \hline
	\end{tabular}
 }
\end{table}
\begin{table}[hbtp]
	\centering
	\setlength{\tabcolsep}{1.8pt}
	\caption{Number of Selected Rules based on Criteria for different $n$ from $162$ Rules of Table~\ref{tab:revRules}. }\label{tab:criteria23_rulesize}
	\begin{tabular}{|c|c|c|c|c|c|c|c|c|}
		\hline
		$\mathbf{n} \rightarrow$  & \textbf{6} & \textbf{7} & \textbf{8} & \textbf{9} & \textbf{10} & \textbf{11} & \textbf{12 }& \textbf{13 }\\
		\hline
		$l_1 \rightarrow$ & 0.6 & 0.5 & 0.4 & 0.4 & 0.4 & 0.4 & 0.4 & 0.4 \\
		\hline
		Criteria 2 Rules & 34 & 44 & 44 & 50 & 44 & 47 & 50 & 45\\
		\hline
		$l_2 \rightarrow$ & 2 & 4 & 6 & 10 & 15 & 20 & 30 & 40 \\
		\hline
		Criteria 3 Rules & 6* & 9* & 15 & 11 & 7 & 5 & 7 & 4\\
		\hline
		Total Candidate Rules & 34 & 44 & 38 & 54 & 48 & 51 & 52 & 48\\
		\hline
	\end{tabular}
\end{table}

Here also, the maximum allowed number of cycles ($l_2$) can be chosen based on user requirements. Here, for $n=13$, we fix $l_2=40$ and get $4$ such CA rules following Criteria 3. These rules are also shown in Table~\ref{tab:clustering_ex}. Table~\ref{tab:criteria23_rulesize} records the value of $l_2$ chosen in our work for each $n$ and the number of rules we can get for that $l_2$. In this table, for $n=6, 7$, a number of rules are marked with * because, as $n$ is very small, there are only a few configurations. For example, for $n=6$, the configurations are $000000(0)$ to $111111(63)$. So, the hamming distance between configurations will always be two digits. Hence, we take only CAs with a number of cycles = 2 such that all cycles maintain average hamming distance $\le 9$. We get $6$ such rules for $n=6$: $252702735$, $1263225675$, $3789677025$ and $4042321935$ with an average hamming distance of the cycles as $0$ and $260960271$ and $756019215$ where an average hamming distance of both the cycles is $9$. Similarly, for $n=7$, restricting the CA to have only four cycles we get only $9$ rules $252695055, 252702735, 1263225675$, $3035673735$, $3785744805$, $3789677025$, $4041289185$, $4042310415$ and $4042321935$ which maintain average hamming distance $\le 9$ for $50\%$ of the cycles.

\begin{table}[!ht]
	\centering
	\caption{List of Rules satisfying Criteria 2 and 3 for $n=6$ to $12$}\label{tab:rulelist}
	\resizebox{0.9\textwidth}{6cm}{
		\begin{tabular}{|l|l|l|l||l|l|l|l|}
			\hline
			\multicolumn{4}{|c||}{$n=12$}& 	\multicolumn{4}{c|}{$n=10$} \\ \hline
			1259293455 & 252691440 & 265482450 & 4035440880 & 1259293455 & 252691215 & 263458635 & 4034007024 \\ \hline
			1263225615 & 252695055 & 267390735 & 4039315215 & 1263225615 & 252691440 & 264499440 & 4035440880 \\ \hline
			1263225675 & 252698895 & 2783028705 & 4041289185 & 1263225675 & 252695055 & 265482450 & 4041289185 \\ \hline
			1921479288 & 252702735 & 3031741620 & 4042264560 & 1921479288 & 252698895 & 267390735 & 4042264560 \\ \hline
			2018211960 & 254611245 & 3538955760 & 4042272240 & 2018211960 & 252702735 & 2783028705 & 4042272240 \\ \hline
			2018212080 & 254618925 & 3789630945 & 4042310415 & 2018212080 & 254611245 & 3537972705 & 4042310415 \\ \hline
			2273806215 & 255594255 & 3789677025 & 4042321935 & 2273806215 & 254618925 & 3538955760 & 4042321935 \\ \hline
			252645195 & 255652080 & 4027518735 & 505290270 & 252645195 & 255594255 & 4027518735 & 505290270 \\ \hline
			252645360 & 256577355 & 4027544304 & 517140690 & 252645360 & 255652080 & 4027544304 & 517140690 \\ \hline
			252648975 & 259526415 & 4027576560 & 521018910 & 252648975 & 256577355 & 4027576560 & 521018910 \\ \hline
			252656655 & 260509455 & 4030467855 & 755961615 & 252656655 & 259526415 & 4030467855 & 756019215 \\ \hline
			252656880 & 264441615 & 4031508720 & 755969295 & 252656880 & 263458575 & 4031508720 & 757935405 \\ \hline\hhline{~~~~----}
			252691215 & 264499440 & 4034007024 & 757935405 & \multicolumn{4}{c|}{~} \\ \hhline{----~~~~}\hhline{----~~~~}
			\multicolumn{4}{|c||}{~}& \multicolumn{4}{c|}{$n=9$}\\\hhline{~~~~----}
			\multicolumn{4}{|c||}{$n=11$}& 1259293560 & 252695055 & 3027809460 & 4042264560 \\ \hline
			1259293455 & 252691440 & 265482450 & 4039315215 & 1263225615 & 252698895 & 3538955760 & 4042268400 \\ \hline
			1263225615 & 252695055 & 267390735 & 4041289185 & 1263225675 & 252702735 & 3726716205 & 4042272240 \\ \hline
			1263225675 & 252698895 & 2783028705 & 4042264560 & 1921479288 & 254618925 & 3789630945 & 4042310415 \\ \hline
			1921479288 & 252702735 & 3031741620 & 4042272240 & 2018211960 & 255652080 & 3789677025 & 4042321935 \\ \hline
			2018211960 & 254611245 & 3538955760 & 4042310415 & 2018212080 & 256577355 & 4027518735 & 505290270 \\ \hline
			2018212080 & 254618925 & 3789677025 & 4042321935 & 2273806215 & 259526415 & 4027544304 & 517140690 \\ \hline
			2273806215 & 255594255 & 4027518735 & 505290270 & 2495001780 & 260960271 & 4027576560 & 521018910 \\ \hline
			252645195 & 255652080 & 4027544304 & 517140690 & 252645195 & 263458575 & 4030467855 & 570174960 \\ \hline
			252645360 & 256577355 & 4027576560 & 521018910 & 252645360 & 263458635 & 4031508720 & 756019215 \\ \hline
			252648975 & 259526415 & 4030467855 & 755961615 & 252648975 & 264499440 & 4034007024 & 757935405 \\ \hline
			252656655 & 260509455 & 4031508720 & 756011535 & 252656655 & 265482450 & 4035440880 & 764273037 \\ \hline
			252656880 & 260960271 & 4034007024 & 757935405 & 252656880 & 267390735 & 4039315215 & ~ \\ \hline
			252691215 & 264499440 & 4035440880 & ~ & 252691440 & 267422991 & 4041289185 & ~ \\ \hline \hline
			\multicolumn{4}{|c||}{~} & \multicolumn{4}{c|}{~}\\ 
			\multicolumn{4}{|c||}{$n=8$} & \multicolumn{4}{c|}{$n=7$} \\ \hline
			1259293515 & 2373488007 & 4027518735 & 4042264560 & 1259293515 & 252702735 & 4027544304 & 4042272240 \\ \hline
			1263225615 & 2495001780 & 4027544304 & 4042272240 & 1263225675 & 260960271 & 4027576560 & 4042275855 \\ \hline
			1267157880 & 252702735 & 4027576560 & 4042310415 & 1267157835 & 3035673735 & 4030467855 & 4042310415 \\ \hline
			1921479288 & 260960271 & 4029484845 & 4042321935 & 1799965515 & 3278049123 & 4031508720 & 4042321935 \\ \hline
			2018211960 & 3027809460 & 4030467855 & 756011535 & 2072267595 & 3373255107 & 4034007024 & 756011535 \\ \hline
			2018212080 & 3063191190 & 4031508720 & 756019215 & 2076199755 & 3726716205 & 4035440880 & 756019215 \\ \hline
			2218767375 & 3538955760 & 4034007024 & 757935405 & 2218767375 & 3777830445 & 4039315215 & 764269197 \\ \hline
			2273806215 & 3777826605 & 4035440880 & 764269197 & 2276755275 & 3785744805 & 4039373040 & 764276877 \\ \hline
			2275772325 & 3789630945 & 4039315215 & ~ & 2373488007 & 3789677025 & 4041289185 & ~ \\ \hline
			2276755335 & 3789677025 & 4041289185 & ~ & 252695055 & 4027518735 & 4042264560 & ~ \\ \hline\hline
			\multicolumn{8}{|c|}{~} \\ \hline
			\multicolumn{8}{{c}}{\begin{tabular}{|l|l|l|l|l|l|l|}
					\multicolumn{7}{|c|}{$n=6$}  \\ \hline
					1263225675 & 2018212080 & 252695055 & 267390735 & 4027544304 & 4035440880 & 4042321935\\ \hline
					1511938590 & 252645336 & 252702735 & 267422991 & 4027576560 & 4039315215 & 570174960 \\ \hline
					1831415085 & 252645360 & 255652080 & 3538955760 & 4030467855 & 4042264560 & 756011535  \\ \hline
					1921479288 & 252656880 & 260960271 & 3789677025 & 4031508720 & 4042272240 & 756019215\\ \hline
					2018211960 & 252691440 & 264499440 & 4027518735 & 4034007024 & 4042310415 & \\ \hline
			\end{tabular}}
	\end{tabular}
 }
\end{table}

Our final list of CA rules is the union of the set of rules selected following criteria 2 and 3. These CAs are good candidates for effective clustering having a good mix of rules with very low hamming distance and a limited number of cycles. The number of such rules is shown in the last row of Table~\ref{tab:criteria23_rulesize}. For example, for $n=13$, there are $48$ such unique rules (the rule $252691440$ is repeated in both criteria). The detailed list of selected rules for each $n=\{6,\cdots,12\}$ is shown in Table~\ref{tab:rulelist}. With these rules, we can proceed to our clustering algorithm where we use only these rules for clustering with binary CA.

\section{Hierarchical Clustering Algorithm with Reversible CA}
As mentioned already, the problem with reversible CA is that for any cell length $(n)$ and any dataset, we may not get the perfect bijective global transition function that can cluster the dataset into the desired number of clusters based on its configuration space. So for this, we need to do this clustering hierarchically.
High-dimensional dataset clustering using the cyclic space of reversible finite cellular automata can be done in three stages. But before applying a CA rule, the dataset needs to be converted into binary. Here, as part of this preprocessing stage, all the dataset objects can be encoded based on frequency encoding similar to the same shown in Table \ref{tab:clustering_ex}. 

\begin{figure}
	\centering
	\includegraphics[width=1.1\linewidth]{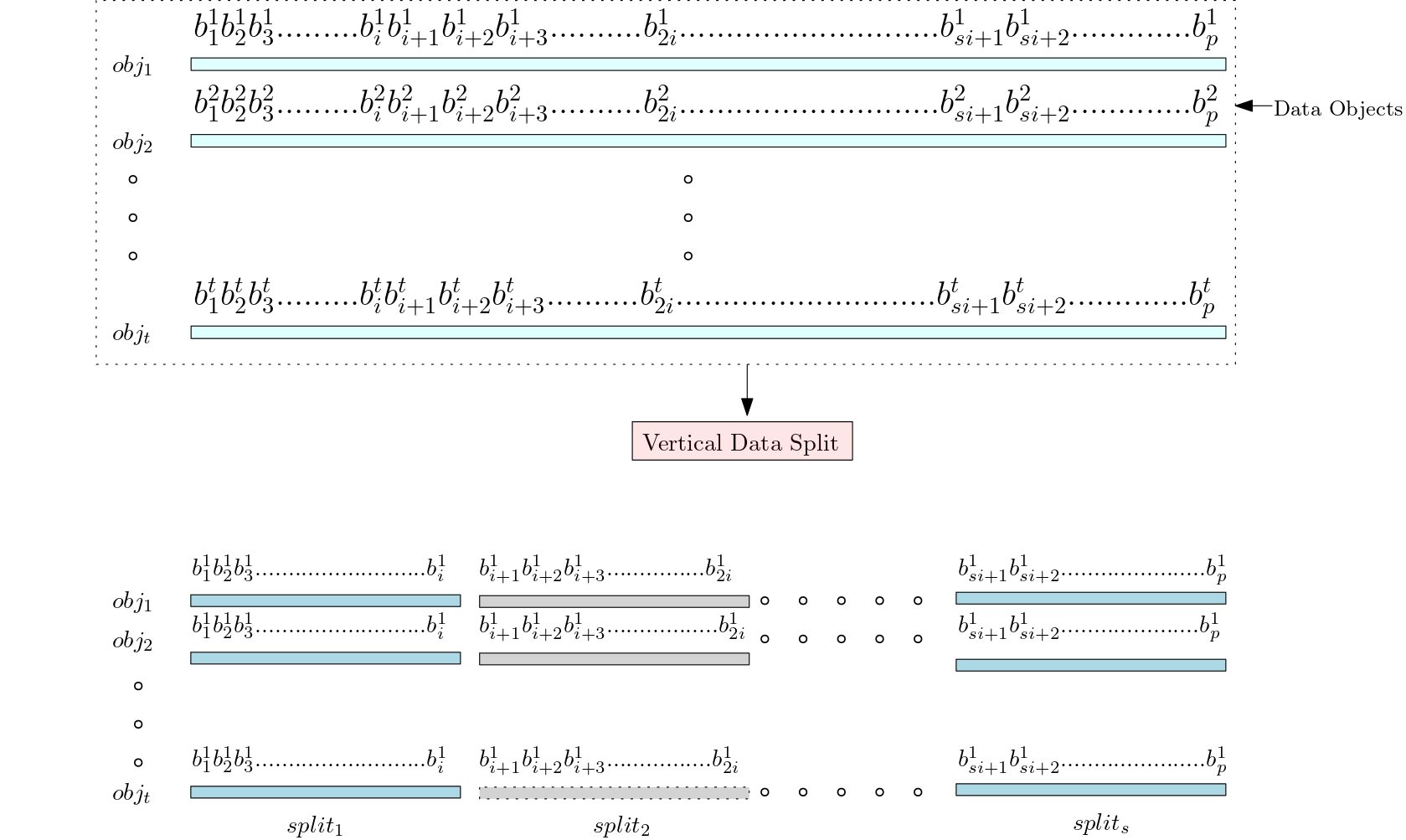}
	\caption{Vertical data split}
	\label{fig:Vertical_split}
\end{figure}
\par After frequency-based encoding, the dataset will be in the form of a set of binary strings. Each binary string corresponds to each object in the dataset. Let $p$ denote the length of each binary string and $t$ denote the total number of binary strings in the dataset. For example, in Table \ref{tab:clustering_ex}, the length of each object in binary form is $5$, and the total number of objects is $8$. That means in the hypothetical dataset shown in the Table \ref{tab:clustering_ex}, $p$ equals $5$ and $t$ equals $8$. Let  $\mathbb{X}=\{X_1, X_2,\cdots, X_{t}\}$  represent the set of all dataset elements. These dataset elements can be encoded using frequency-based encoding, and each object can be mapped to a target configuration, represented as a binary string. Let $F_D =\{obj_1,obj_2,\cdots,obj_i,\cdots,obj_t\}$ represent the set of $t$ binary strings that correspond to each object in the dataset. Example \ref{EXample:School_District_frequency_encode} illustrates the use of frequency-based encoded objects with a real dataset.

\begin{example}
	Consider the School District Breakdowns data set from the Data world repository where dataset elements can be represented by the set $\mathbb{X}=\{X_1, X_2,\cdots, X_{32}\};$ that is $\lvert\mathbb{X} \rvert = 32$. There are 44 attributes, and each tuple can be encoded using the frequency-based encoding (mentioned in Table:\ref{tab:clustering_ex}), each target object is mapped to a target configuration, represented as a set of binary strings ($F_D$). Here $F_D =\{obj_1,obj_2,\cdots,obj_i,\cdots,obj_{32}\}$, where,\\
	$obj_1$ =
	010101011000000101000101000011111111000000000\\0000111000001100000000001010000011000000101, \newline
	$obj_2$ =
	111101101100001101001011000011110101111111010\\0001111101110110000000011011011101100
	001101\newline and so on till $obj_{32}$.
	\label{EXample:School_District_frequency_encode}
\end{example}

Our hierarchical clustering algorithm will take this set of strings $F_D =\{obj_1, obj_2,\cdots,obj_i,\cdots,obj_t\}$, as input and do clustering using three stages.

\subsection{Stage 1 - Initial Clustering}

The dataset consists of $t$ number of objects, where each object is a $p$ bit binary string. Let $F_D =\{obj_1,obj_2,\cdots ,obj_j,\cdots ,obj_t\}$ be the set of $t$ objects where $obj_j= b_1^jb_2^j b_3^j \cdots b_i^j\cdots b_p^j$. Here $b_i^j$ represents $i^{th}$ bit of $j^{th}$ object. In a low-dimensional dataset, the length $p$ of each object will be small. For such cases, we can take this $p$-bit string directly as the configuration of our CA. But in the case of the high-dimensional dataset, the length $p$ will be large. To reduce computational costs, we must split each object into many splits and then take each split into separate configurations. As shown in Figure \ref{fig:Vertical_split}, each object is split into $s$ configurations such that $obj_j = \langle split_1|split_2|\cdots|split_s \rangle$ where the string `$b_1^jb_2^jb_3^j \cdots b_i^j$' $\in$ $split_1$, `$b_{i+1}^jb_{i+2}^jb_{i+3}^j\cdots b_{2i}^j$' $\in split_2$, $\cdots$,`$b_{si+1}^jb_{si+2}^j \cdots b_p^j$' $\in$ $split_s$. The size of each split is equal to the cell length of the CA we choose, and the number of splits depends on the split size. That means $\lfloor s=\dfrac{p}{n_1} \rfloor$, where $s$ is the number of splits, $p$ is the length of the object in binary string format, and $n_1$ is the split size. Let $x\restriction_{split_i}$ represent the set of strings in the $i_{th}$ split of all objects. If there are $t$ objects, then $\lvert x\restriction_{split_i} \rvert$  is $t$.

\begin{figure}
	\centering
	\includegraphics[width=1\linewidth]{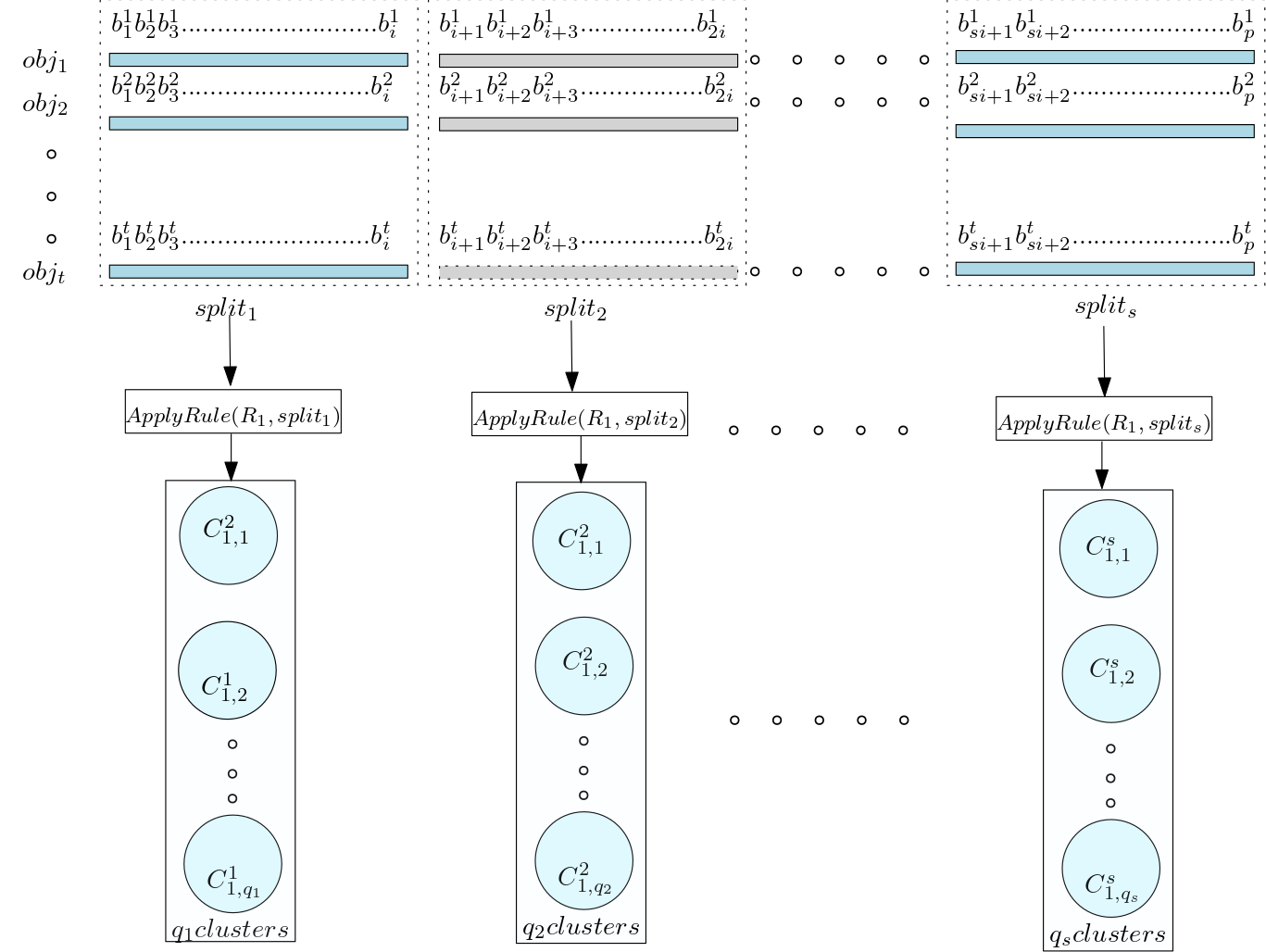}
	\caption{Stage 1 - Initial clustering by applying rule $R_1$ to each vertical split.}
	\label{Fig:Stage 1}
\end{figure}
\begin{example}

Consider the school district breakdown dataset with $44$ attributes. The frequency-based encoded binary string in Example \ref{EXample:School_District_frequency_encode} for $obj_1$ can be split into $7$ splits if we take split size as 13: $split_1$= `0101010110000', $split_2$ = `0010100010100', $split_3$ = `0011111111000', $split_4$ = `0000000000111', $split_5$ = `0000011000000', $split_6$ = `0000101000001', $split_7$ = `1000000101'. Similarly $obj_2$ can be split as: $split_1$= `1111011011000', $split_2$ = `0110100101100', $split_3$ = `0011110101111', $split_4$ = `1110100001111', $split_5$ = `1011101100000', $split_6$ = `0001101101110', $split_7$ = `1100001101' and so on till $obj_{32}$.  Then \newline$x\restriction_{split_1}$ =\{`0101010110000',`1111011011000', $\cdots\}$,
	\newline$x\restriction_{split_2}$ =\{`0010100010100',`0110100101100', $\cdots\}$,
	\newline$x\restriction_{split_3}$ =\{`0011111111000',`0011110101111', $\cdots\}$,
	\newline$x\restriction_{split_4}$ =\{`0000000000111',`1110100001111', $\cdots\}$,
	\newline$x\restriction_{split_5}$ =\{`0000011000000',`1011101100000', $\cdots\}$,
	\newline$x\restriction_{split_6}$ =\{`0000101000001',`0001101101110', $\cdots\}$,
	\newline$x\restriction_{split_7}$ =\{`1000000101', `1100001101', $\cdots\}$, each $x\restriction_{split_i}$ contains $32$ strings each of length $13$ except the last split which may contain strings of smaller length.
	\label{EX:School_datset_split}
\end{example}
\begin{algorithm}[H]
	\begin{algorithmic}[1]
		\REQUIRE $F_D =\{obj_1,obj_2,\cdots,obj_i,\cdots,obj_t\}$ where $obj_i = \langle split_1,split_2,\cdots,split_s \rangle$ 
		\ENSURE Clusters=$\{\{C^1_{1,1},C^1_{1,2},\cdots,C^1_{1,q_1}\},\{C^2_{1,1},C^2_{1,2},\cdots,C^2_{1,q_2}\},\cdots,$
       $\{C^s_{1,1},C^s_{1,2},\cdots,C^s_{1,q_s}\}\}$
		\STATE $Clusters = \{\}$
		\FOR{each  i from 1 to s}
		\STATE $Cluster_i\gets ApplyRule(R_1,x\restriction_{split_i})$
		\STATE $Clusters = Clusters \cup cluster_i$
		\ENDFOR
	\end{algorithmic}
	\caption{Stage 1 - Initial clustering by applying rule $R_1$ to each vertical split.}
	\label{alg:stage1}
\end{algorithm}
After doing the vertical split, the rule $R_1$ is applied for all  $x\restriction_{split_i}$. After applying rule $R_1$, $x\restriction_{split_1}$ is grouped into $q_1$ clusters, $x\restriction_{split_2}$ into $q_2$ clusters, and $x\restriction_{split_s}$ is grouped into $q_s$ clusters as shown in Figure \ref{Fig:Stage 1}. Let $C^{split}_{stage, cycle}$ represents the cycles of each split in different stages. The clusters generated during Stage 1 are $ \{C^1_{1,1},C^1_{1,2}, \cdots ,C^1_{1,q_1}\}$, $\{C^2_{1,1},C^2_{1,2}, \cdots ,C^2_{1,q_2}\}$,$\cdots$,  $C^s_{1,1},C^s_{1,2}, \cdots ,C^s_{1,q_s}\}\}$. The generation of clusters in Stage 1 is shown in Algorithm \ref{alg:stage1}. Over these clusters, we now apply Stage 2 to merge them. In summary, Stage 1 comprises the following steps:
\begin{description}
	\item \textbf{Step 1:} Split the frequency-based encoded objects in the dataset.
	\item \textbf{Step 2:} Apply the CA rule to each split and generate an initial set of clusters.
\end{description}

\subsection{Stage 2 - Median-based clustering with a new rule}
The clusters formed during Stage 1 consist of many cycles where parts of the same object belong to different clusters. We need to merge these splits in such a way that, after merging, the length of the new configuration for the same data object is reduced. This is to ensure that the computational cost of clustering all elements is much reduced. To do this, we take an indexing approach. We first sort the cycle in each split in the increasing order of the median of the elements in the cycle. Two cycles with median values close to each other indicate these cycles' elements are closer than some other cycles for which the difference between the medians is high. Now, we number each cycle based on its position in the sorted order and represent that number in Gray code. Gray code is chosen because, in the Gray code representation of binary strings, the hamming distance between two consecutive strings is less. Then replace elements in each split by the cycle number to which the element belongs. Let $Clusters = \{\{C^1_{1,1},C^1_{1,2},\cdots,C^1_{1,q_1}\},\{C^2_{1,1},C^2_{1,2},\cdots,C^2_{1,q_2}\},\cdots,\{C^s_{1,1},C^s_{1,2},\cdots,C^s_{1,q_s}\}\}$ be the clusters formed in Stage 1 and  $q_j clusters=\{C^j_{1,1}, C^j_{1,2},\cdots,C^j_{1,k},\cdots, C^j_{1,q_j}\}$ be the cycles generated from $x\restriction_{split_j}$. We first sort these cycles based on the median of elements in each cycle. After sorting if $split_q \in C^q_{1,t}$ where $split_q$ is the configuration belonging to the $t^{th}$ cycle, then change the value of $split_q$ into a Gray code value of $t$. Example \ref{example: Changing to gray code} shows these steps using a hypothetical dataset.
\begin{algorithm}[H]
	\begin{algorithmic}[1]
		\REQUIRE $Clusters = \{\{C^1_{1,1},C^1_{1,2},\cdots, C^1_{1,q_1}\},\{C^2_{1,1},C^2_{1,2},\cdots,C^2_{1,q_2}\},\cdots,$\newline $\{C^s_{1,1}, C^s_{1,2},\cdots,C^s_{1,q_s}\}\}$\newline
		$F_D =\{obj_1,obj_2,\cdots,obj_i,\cdots,obj_t\}$ where $obj_i = \langle split_1|split_2|\cdots |split_s \rangle$ 
		\ENSURE $Clusters = \{C^1_{2,1},C^1_{2,2},\cdots,C^1_{2,u}\} $
		$F_D =\{obj_1,obj_2,\cdots,obj_i,\cdots,obj_t\}$ where $1 \le obj_i \le u$ ; $u = $Number of cycles
		\FOR{each  clusters $i = 1$ to $s$} 
		\FOR{each  cycle $j$}
		\STATE {$Median_j$ $\gets$ Find Median($cycle_j$)}  /*\textit{Find the median of elements in the cycles in each cluster}*/
		\STATE $Median = Median$ $\cup$ $Median_j$
		\ENDFOR
		\STATE {Sort $cycles$ based on $Median$}
		\FOR{Each  $split_q$ in $F_D$}
		\IF{$ split_q \in cycle_k$}
		\STATE $split_q \gets Gray code(k)$ /*\textit{change the value of $split_q$ into  Gray code of cycle index. }*/
		\ENDIF
		\ENDFOR
		\ENDFOR
		\FOR{each  $obj_j \in F_D$}
		\STATE $obj_j \gets g_1^jg_2^jg_3^j \cdots g_{k_1}^j|$$g_{{k_1}+1}^jg_{{k_1}+2}^jg_{{k_1}+3}^jg_{k_2}^j|\cdots $\newline$|g_{{k_{s-1}}+1}^jg_{{k_{s-1}}+2}^jg_{{k_{s-1}}+3}^j\cdots g_{k_s}^j$ /*\textit{Merge all split of each object}*/
		\ENDFOR
		\STATE {$length_x$ $\gets$ length of $obj_i \in F_D$} /*\textit{Find length of object after merging}*/
		\IF{$length_x \le  $ Maximum allowed cell length}
		\STATE $cluster_l\gets ApplyRule(R_2,F_D)$
		\STATE Median = $\{\}$
		\FOR{each  cycle $j$ $\in$ $cluster_l$}
		\STATE {$Median_j$ $\gets$ Find Median($cycle_j$)}
		\STATE Median = Median $\cup$ $Median_j$
		\ENDFOR
		\STATE {Sort $cycles$ based on $Median$}
		\FOR{Each  $obj_q$ $\in$ $F_D$}
		\IF{$ obj_q \in cycle_k$}
		\STATE $obj_q \gets k$ /*\textit{Change object value with cycle index to which it belongs}*/
		\ENDIF
		\ENDFOR
		\ELSE
		\STATE CALL $Stage1$ and $Stage2$ using Rule $R3$
		\ENDIF
	\end{algorithmic}
	\caption{Stage 2 - Clustering using the median of each cycle and applying Rule $R_2$}
	\label{alg:stage2}
\end{algorithm}

\begin{example}
	Consider the hypothetical dataset explained in Table \ref{tab:clustering_ex}. Now let us use the CA $26422991$ of Figure \ref{Fig:cycleEx} to cluster these objects. This hypothetical dataset has less number of dimensions, so only one split is required. Here $x\restriction_{split_1}$ =$\{$`00001', `01001', `00010', `00100', `01100', `01100', `11001', `11010'$\}$. Now apply Rule $R1=267422991$ on  $x\restriction_{split_1}$. By applying rule $R_1$, the $x\restriction_{split_1}$ will be grouped into $3$ cycles as shown in Figure \ref{Fig:cycleEx}. Then sort the cycle based on the median of elements in each cycle. Next, we need to replace elements in each cycle using the cycle index to which it belongs. Table \ref{tab: Stage2_Cycle_gray} shows the object's initial configuration and new configuration based on Gray code encoding of the cycle index to which the element belongs. For example, $Obj_1 \in Cycle_0$ then $Obj_1$ new configuration is $00 ($Gray code of cycle index$=0)$ similarly for remaining elements in $x\restriction_{split_1}$. The updated $x\restriction_{split_1}$ = \{`00', `00', `11', `01', `00', `00', `01', `01'\}.
	
	\label{example: Changing to gray code}
\end{example}

\begin{figure}
	\centering
	\includegraphics[width=1\linewidth]{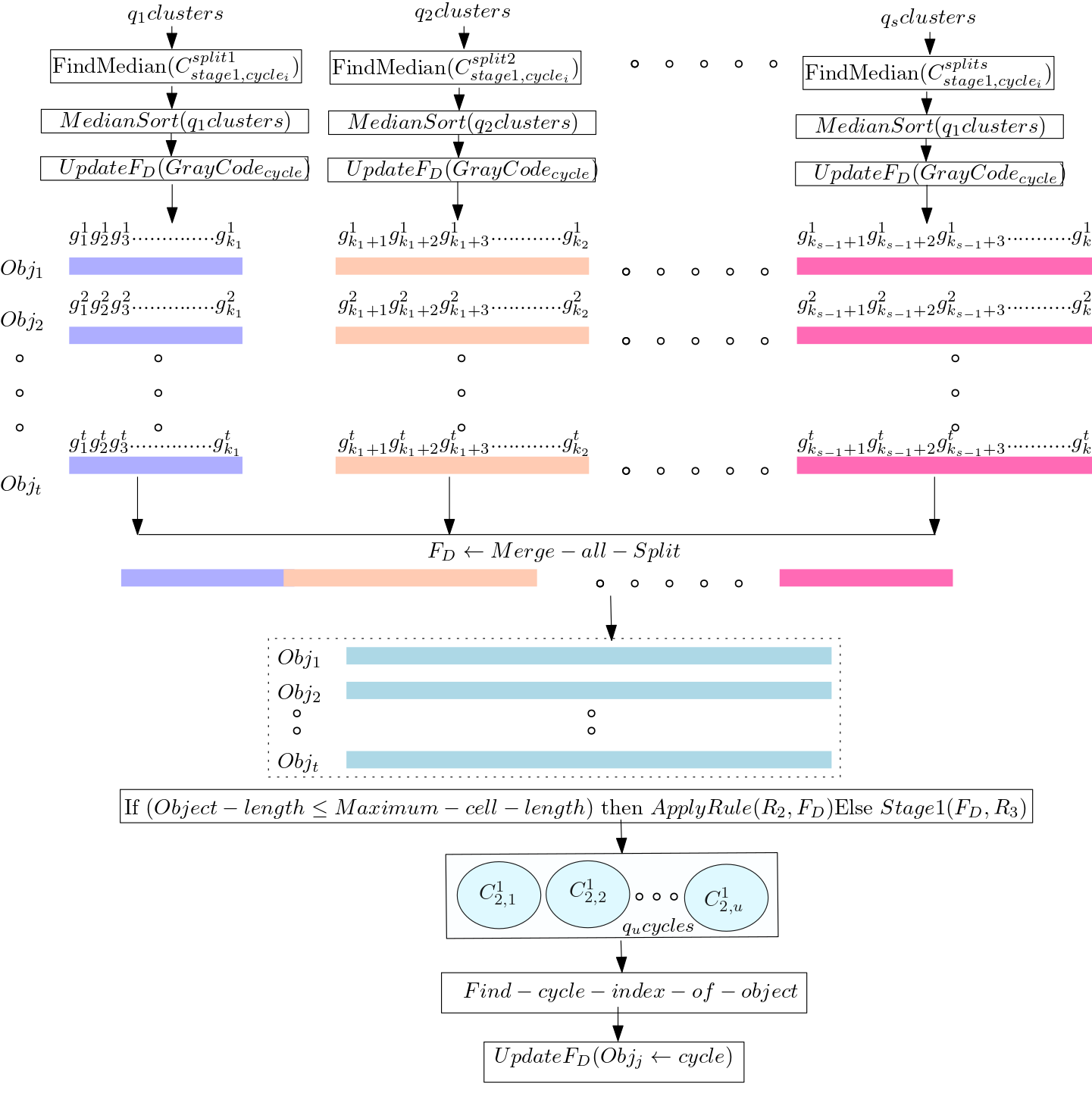}
	\caption{Stage 2 - Clustering using the median of each cycle and applying rule $R_2$}
	\label{fig: Stage 2}
\end{figure}

\begin{table}[h]
	\setlength{\tabcolsep}{1.4pt}
	\centering
	\caption{Stage2: Next level encoding of objects}
	\label{tab: Stage2_Cycle_gray}
			\resizebox{0.7\textwidth}{!}{
		\begin{tabular}{|c|c|c|c|}
			\hline
			\multirow{2}{*}{Objects} & {Initial Configuration} &{Cycle Index}&{New configuration}\\
			
			& (with Decimal Equivalent) &{to which object belongs}&{(in Gray code)}\\
			\hline
			$Obj_1$ &  00001 (1)&Cycle 0 & 00\\
			\hline
			$Obj_2$ &  01001 (9)&Cycle 0 & 00\\
			\hline
			$Obj_3$ &  00010 (2)&Cycle 2 & 11\\
			\hline
			$Obj_4$ & 00100 (4)&Cycle 1 & 01\\
			\hline
			$Obj_5$ &  01100 (12) &Cycle 0 & 00\\
			\hline
			$Obj_6$ & 01100 (12) &Cycle 0& 00\\
			\hline
			$Obj_7$ &  11001 (25) &Cycle 1&01\\
			\hline
			$Obj_8$ & 11010 (26) &Cycle 1&01\\
			\hline
		\end{tabular}
			}
\end{table}

In the Example \ref{example: Changing to gray code} of a hypothetical dataset, there is only one split. But in a real dataset, the dimension will be high, therefore we require more splits, as shown in the Example \ref{EX:School_datset_split} of the School District Breakdowns data set, which requires 7 splits if the split size is 13. In the case of a high-dimensional dataset, we encode elements in each split using the same concept shown in Table \ref{tab: Stage2_Cycle_gray}. Then we concatenate all the splits to get a single binary string of objects. The initial configuration of
$F_D =\{obj_1,obj_2,\cdots \cdots,obj_i,\cdots,obj_t\}$ where 
$obj_j = \langle split_1|split_2|split_3|\cdots \cdots|split_s \rangle,$ where 
`$b_1^jb_2^jb_3^j \cdots\cdots b_i^j$' $\in split_1$,
`$b_{i+1}^jb_{i+2}^jb_{i+3}^j \cdots b_{2i}^j$' $\in split_2$
,`$b_{si+1}^jb_{si+2}^j \cdots b_p^j$' $\in$ $split_s$. 
Let the updated $F_D =\{obj_1,obj_2,\cdots \cdots,obj_i,\cdots,obj_t\}$ where
$obj_j = \langle split_1' |split_2'|split_3'|\cdots \cdots |split_s' \rangle$, 
$g_1^jg_2^jg_3^j \cdots g_{k_1}^j \in split_1'$, $g_{{k_1}+1}^j g_{{k_1}+2}^jg_{{k_1}+3}^j \cdots g_{k_2}^j\in split_2'$, $g_{{k_{s-1}}+1}^jg_{{k_{s-1}}+2}^jg_{{k_{s-1}}+3}^j\cdots g_{k_s}^j$ $\in$ $split_s$. 

Here $split_i$ is reduced to $split_i'$, where each object of $split_i$. clustered into 
$\{C^i_{1,1}, C^i_{1,2},\cdots, C^i_{1,{q_i}}\} $ is represented by the Gray code corresponding to $t$ where $t$ is the cycle number to which the part of the object belongs, $1\le t \le q_i$. So $\lvert{split_1'}\rvert = k_i$ where $k_i$ is the number of bits required to represent $q_i$ as shown in Figure \ref{fig: Stage 2}. After this update, merge all the splits into one single binary string. Then $F_D =\{obj_1,obj_2,\cdots,obj_i,\cdots,obj_t\}$ will contain $obj_1$, $obj_2$,$\cdots$,$obj_t$ without any splits. Let $n_2$ represent the length of the object in the dataset $F_D$ after merging then $n_2$ is $k_1$+$k_2$+$\cdots$+$k_s$.

\par If the length of all objects $obj_i \in F_D$ is less than the maximum allowed cell length $(n)$, Then apply rule $R_2$ to the dataset $F_D$. Otherwise, repeat Stages 1 and 2 using rule $R_3$ until the length of all object $obj_i \in F_D$ is less than the maximum possible cell length. This value of $n$ depends on the computational power of the user's system and is as per the user's choice. When the cell length is n, there will be a $2^n$ configuration. If our system is capable of handling $2^n$ configuration efficiently, there is no need for a recursion call; otherwise, we need to reduce the length of the object by doing states 1 and 2 recursively. Then by applying the rule $R2$ on the objects in the dataset $F_D$, it will grouped into different cycles $C^1_{2,1},C^1_{2,2},\cdots,C^1_{2,u}$  as shown in Figure \ref{fig: Stage 2}. So, if the computational power of the system is high, one can directly apply rule $R2$ without calling stages 1 and 2 recursively.

\par Still, the number of clusters may be larger than the desired number of clusters. So, we need to merge these clusters. For this, we again take the median-based approach to find the closeness of elements present in different cycles, we first need to find the median of elements in the cycle. If the median of elements in the cycle has less difference, it means they are related. So by sorting the cycles based on the median of elements in the cycle, the inter-cycle relationship can be maintained. The newly created cycles after sorting can be considered as new clusters. Let $Clusters = \{C^1_{2,1},C^1_{2,2},\cdots,C^1_{2,u}\} $ be the clusters generated in Stage 2. The number of cycles after Stage 2 is $u$. Each object will belong to only one of these $u$ cycles. Now, update $F_D =\{obj_1,obj_2,\cdots,obj_i,\cdots,obj_t\}$ with the cycle index to which the object belongs. If $obj_q \in C^1_{2,u}$, then change the $obj_q$ into $u$. 

In summary,  Stage 2 involves the following steps (see Algorithm~\ref{alg:stage2}):
\begin{description}
	\item \textbf{Step 1:} Find the median of elements in each cycle, sort the cycles in each split based on the median, and index the cycles based on the sorting.
	\item \textbf{Step 2:} Update elements in each split into the Gray code representation of the cycle index to which the element belongs.
	\item \textbf{Step 3:} Merge elements in all split to get a single binary string representation of objects in the dataset.
	\item \textbf{Step 4:} If the length of the merged object is less than the maximum possible cell length, then apply a new rule to the single split dataset and generate a new set of cycles; otherwise, repeat Stage 1.
 \item \textbf{Step 5:}Sort the new cycles based on the median of the cycle elements and index the cycles.
	\item \textbf{Step 6:} Update the dataset by replacing the object with the cycle index to which it belongs.
	
\end{description}

\subsection{Stage 3 - Clustering using cycle median gap}
\begin{figure}
	\centering
	\includegraphics[width=.6\linewidth]{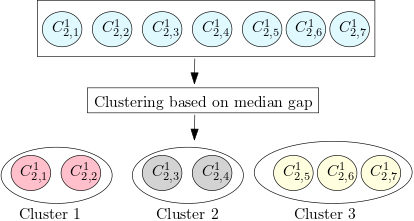}
	\caption{Example for Stage 3: Three clusters are created based on two maximum median gaps at the cycle index at 2 and 4}
	\label{fig:stage3-ex}
\end{figure}

The final clustering can be done by using the median gap. Median gap means the difference between the median of elements in the cycle. When the difference between the median of elements in two cycles is high, the inter-cycle relationship between those two cycles is less that is these elements are unrelated. However, if the cycles' median values are very close, that means the elements of these cycles are closely related, and they can be clustered into the same cluster, maintaining the minimum intra-cluster distance. To do this in Stage 3, we find the difference between the medians of the elements of cycles. If we need $k_c$ clusters, we need to find the $k_c - 1$ median gap. The maximum median gap between the median of elements in the cycle can be used as the final constraint for creating the cluster. Stage 3 consists of the following steps:
\begin{description}
	\item \textbf{Step 1:} If we want to generate $k_c$ number of clusters, then find $k_c-1$ median gaps between the cycles generated in Stage 2.
	\item \textbf{Step 2:} Group the cycles into $k_c$ clusters based on median gaps.
\end{description}
\begin{algorithm}[H]
	\begin{algorithmic}[1]
		\REQUIRE$Clusters = \{C^1_{2,1},C^1_{2,2},\cdots,C^1_{2,u}\} $
		$F_D =\{obj_1,obj_2,\cdots,obj_i,\cdots,obj_t\}$\newline where $1 \le obj_i \le u$
		\ENSURE $k_c$ clusters
		
		\FOR{$i=1$ to $k_c - 1$}
		\STATE $P_i$ $\gets$ Index of $i^{th}$ Maximum-difference-in-Median /*\textit{Obtain $k_c - 1$ gap position}*/
		\ENDFOR
		\STATE Sort P(Index of gap)
		\FOR{Each  $obj_q$ $\in$ $F_D$}
		\STATE $obj_q \gets $ cluster number /*\textit{Replace object with cluster number based on cycle index and Median gap }*/
		\ENDFOR
	\end{algorithmic}
	\caption{Stage 3 - Final clustering using cycle median gap}

	\label{alg:stage3}
\end{algorithm}
Figure \ref{fig:stage3-ex} shows the clustering of elements into three clusters using two maximum median gaps at cycles 2 and 4. The maximum median gap at cycle index 2 represents a large median gap between cycles 2 and 3, which means there is less inter-cluster relationship between cycles 2 and 3. Similarly, the maximum median gap at cycle index 4 represents a large median gap between cycles 4 and 5, which means there is less inter-cluster relationship between cycles 4 and 5. Then we can cluster these elements as follows: data elements in cycles 1 and 2 belong to Cluster 1, data elements in cycles 3 and 4 belong to Cluster 2, and data elements in cycles 5, 6, and 7 belong to Cluster 3.

\par Let $Clusters = \{C^1_{2,1}, C^1_{2,2},\cdots, C^1_{2,p}, C^1_{2, i}, C^1_{2,j},\cdots, C^1_{2,u}\} $ be the final cycle created in stage 2 and the desired number of clusters is two.  If the maximum difference between the cycle median is for $C^1_{2, i}$ and $C^1_{2,j}$ then clustering of elements can be done using these indices  $i$ and $j$. The set $F_D =\{obj_1,obj_2,\cdots \cdots,obj_h,\cdots,obj_t\}$  can be updated based the cycle index. if $obj_h \in C^1_{2,p}$ where $p \le i$ then $obj_h$ can be grouped into $cluster_1$ else into $cluster_2$ for finding two clusters. These steps are explained in Algorithm \ref{alg:stage3}.

\

		


In Example \ref{Ex:Final clustering of school}, only two clusters are formed. If we need to find $k_c$ clusters we have to find the $k_c - 1$ median gaps and then cluster the elements using the Algorithm \ref{alg:stage3}.

\begin{example}
Consider some hypothetical dataset where $F_D =\{obj_1,obj_2,\cdots,obj_q,\cdots, obj_{32}\}$ and cycles create in Stage 2 is $Clusters =	\{C^1_{2,1}, C^1_{2,2},\cdots,C^1_{2,7}\} $, which means this dataset has 32 objects, and by applying Rule $R_2$ these objects are grouped into seven cycles, $cyle_1, cycle _2, \cdots, cycle_7$. In our proposed algorithm, If $obj_q \in C^1_{2,k}$, then change the $obj_q$ into $k$. For example, in Stage 2 the updated $F_D$ based on cycle to which the data belongs is $F_D =\{1, 3, 1, 1, 4, 1, 1, 1, 1, 4, 3, 2, 1, 1, 1,  3, 1, 1,4, 1, 7, 6, 6, 5, 7, 7, 6, 5, 5, 6, 5, 5\}$, which means $obj_1 \in C^1_{2,1}$, $obj_2 \in C^1_{2,3}$ and so on. To find 2 clusters we need to find one maximum median gap between the cycles. Suppose the maximum median gap is between $cycle_4$ and $cycle_5$. Then objects in cycles 1 to 4 are in $Cluster 1$, and objects in cycles 5 to $7$ will belong to $Cluster 2$. Therefore in Stage 3 we can update $F_D$ as $F_D =\{1, 1, 1, 1, 1, 1, 1, 1, 1, 1, 1, 1, 1, 1, 1, 1, 1, 1, 1, 1, 2, 2, 2, 2, 2, 2, 2, 2, 2, 2, 2, 2\}$. In this example, two clusters are created with the first 20 objects belonging to cluster 1 and the remaining 12 objects belonging to cluster 2.
\label{Ex:Final clustering of school}
\end{example}

\section{Results and Analysis}\label{sec:results}
This section aims to analyze the performance of our algorithm in effectively distributing objects of some benchmark datasets among $k_c$ clusters, where $k_c$ is the desired number of clusters.

Our algorithm requires at least two different rules, one for Stage 1 and one for Stage 2. In Stage 2, for a high-dimensional data set, before applying the second rule, the size of the input object may be greater than the maximum possible cell size. In that situation, we need to repeat Stage 1 using a third rule recursively until the object size becomes smaller than the maximum allowed cell size.

\subsection{Time Complexity }
The proposed clustering algorithm consists of three stages. The worst case happens when in Stage 1 and 2, the number of cycles in each split is equal and each cycle has an equal number of elements. Considering this, the time complexity of each stage can be calculated as follows:
\begin{itemize}
	\item Stage 1: Algorithm\ref{alg:stage1} line number 2 to 5 takes $\mathcal{O}(s* 2^{n_1})$ where $s$ is the number of splits, $n_1$ is split size.
	
	\item Stage 2 - Algorithm\ref{alg:stage2} line number 3  takes $\mathcal{O}(t/k)$ where t is the size of the dataset, and k is the number of cycles (worst case -- number of cycles in each split is equal and each cycle has an equal number of elements). The line number 2 to 5 takes $\mathcal{O}(t/k \times k) = \mathcal{O}(t)$. The line number 6 takes $\mathcal{O}( k \log k)$ for sorting. The line number 7 to 10  takes $\mathcal{O}(t/k \times k) = \mathcal{O}(t)$. Then the total complexity from line number 1 to 12 takes $\mathcal{O}(s \times \max(t,k \log k))$ = $\mathcal{O}(s\times t)$ since the number of cycles ($k$) is very small compared with dataset size ($t$). The line number 13 to 15 takes $\mathcal{O}(t\times s)$. Line number 18 takes $\mathcal{O}(2^{n_2})$ to apply the rule in the dataset where $n_2$ is the updated length of the object in the dataset after merging. The line numbers 20 to 23 take $\mathcal{O}(k\times t/k)= \mathcal{O}(t)$. Line number 24 takes $\mathcal{O}(k \log k)$ for sorting $k$ number of cycles based on the median of elements in the cycle. Lines 25 to 29 take $\mathcal{O}(t)$ time to replace $t$ objects in the dataset with cycle index to with the element belongs. if $n_2$ is large then a recursion call will happen then it takes $t_n \times \mathcal{O} (2^{n_1})$ where $t_n$ is the complexity for a recursive call. Then the overall time complexity of Stage 2 is $\mathcal{O}(2^{n_2})$ when $n_2$ is small. 
	
	\item Stage 3: Algorithm \ref{alg:stage3} line number 1 to 3 takes $\mathcal{O}(k_c)$ where $k_c$ is number of clusters. Line numbers 5 to 7 take $\mathcal{O}(t)$ to replace $t$ number of dataset objects with cluster number based on the median gap. 
\end{itemize}
The overall complexity of our proposed hierarchical clustering algorithm with reversible CA is bounded by $\mathcal{O}(\max(2^{n_1},2^{n_2},s\times t))$. If $n_1$ and $n_2$ are small, which is desirable for faster implementation, our algorithm takes $\mathcal{O}(s\times t)$. Our clustering algorithm drastically reduces the complexity of the clustering technique making it quadratic time. 

On the other hand, if we compare the time complexity with existing state-of-the-art algorithms, time complexity for the k-means algorithm is denoted as $O(KNT)$, where $N$ is the total number of elements in the datasets, $K$ represents the total number of partitions, and $T$ stands for the number of iterations in the clustering process \cite{CL1}. Whereas, for BIRCH algorithm, the time complexity $O(k n^2)$, is contingent on the number of elements to be clustered ($n$) and the specified number of clusters ($k$) \cite{CL1}. Therefore, our proposed clustering algorithm has drastically reduced the complexity of the existing CA based clustering algorithms making it at par with all state-of-the-art algorithms.
\begin{table}[]
	\centering
 \caption{Description of real dataset used for proposed CA-based clustering algorithm}
	\begin{tabular}{ |c|c|c| } 
		\hline
		Dataset & Number of Attributes & Target Objects \\
		\hline
		School District Breakdown & 44 & 32\\ 
		IPL 2018 Data & 24 & 143 \\ 
		Iris & 5 & 150 \\ 
		Buddymove & 6 & 249 \\ 
		heart failure clinical records & 12 & 299 \\ 
		Cervical Cancer Behavior Risk & 19 & 72 \\ 
		seeds & 7 & 210 \\ 
		Wholesale customers data & 6 & 440 \\ 
		StoneFlakes & 7 & 79 \\
            Gas Turbine Emission(2015)&12&7385\\
		\hline
	\end{tabular}
	
	\label{tab:Dataset Description}
\end{table}

\begin{table}[]
	\centering
	\caption{Reversible CA rules used for clustering}
	\resizebox{0.7\textwidth}{!}{
	\begin{tabular}{||c c c c||} 
		\hline
		Dataset & Rule 1 & Rule2 & Rule 3 \\ [0.5ex] 
		\hline\hline
		School District Breakdown & 264499440 & 265482450 & 4042321935 \\ 
		\hline
		IPL 2018 Data & 255652080 & 256577355 & 4041289185\\ \hline
		Iris         & 252691440 & 265482450 & Not required \\ \hline
		Buddymove & 1921479288 &  4041289185 & Not required  \\
		\hline
		heart failure clinical records& 2273806215 & 521018910 & Not required \\ 
		\hline
		Cervical Cancer Behavior Risk & 252645360 & 252648975 & 254618925  \\ 
		\hline
		Seeds & 252691440 & 255652080 & Not required   \\ 
		\hline
		Wholesale customers data & 3031741620 & 757935405  &  Not required \\ 
		\hline
		StoneFlakes & 2018212080 & 2273806215 &  Not required \\ 
		\hline
            Gas Turbine Emission(2015)&1921479288&252691215&4034007024\\
            \hline
	\end{tabular}
 }
	
	\label{tab:Dataset Best result Rules}
\end{table}

\begin{figure}[htbp]
    \centering
   \subfigure[Reversible cellular Automata]{\includegraphics[width=0.4\linewidth]{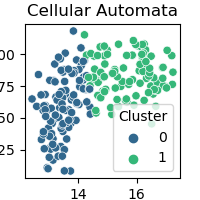}}
    \subfigure[Birch clustering]{\includegraphics[width=0.4\linewidth]{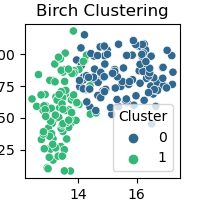}}
    \subfigure[Hierarchical clustering]{\includegraphics[width=0.4\linewidth]{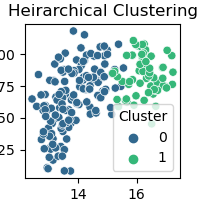}}
    \subfigure[K-means clustering]{\includegraphics[width=0.4\linewidth]{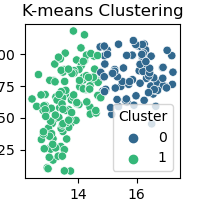}}
    \caption{Illustrating the clustering performed by various algorithms on the seed dataset, represented by two significant attributes: perimeter and asymmetry coefficient.}
    \label{fig:Clustering Diagram}
\end{figure}

\begin{table}[]
	
	\centering
	\caption{Benchmarking algorithms vs scores for Different Datasets }\label{Tab:results}
	\resizebox{0.9\textwidth}{!}{
		\begin{tabular}{|c|c|c|c|c|}
			\hline
			 \bfseries{Dataset} & \bfseries Algorithm  &  \bfseries Silhouette& \bfseries Davis-Bouldin& \bfseries Calinski-Harabasz \\
			 & \bfseries   &  \bfseries Score & \bfseries  Index & \bfseries Index \\
			\hline
			\multirow{5}{*}{\textbf{School District Breakdown}} & \bfseries CA Algorithm  &  \textbf{0.706186}	& \textbf{0.3859} & \textbf{30.6898} \\
			&	\bfseries Hierarchical  & 0.766361 & 0.4342 & 94.4289  \\
			&	\bfseries K-means & 0.766361& 0.4342& 94.4289 \\
			&	\bfseries BIRCH & 0.766361& 0.4342& 94.4289\\
			\hline
			\multirow{5}{*}{\textbf{IPL 2018 Data}} & \bfseries CA Algorithm  &  \textbf{0.434091}	& \textbf{0.3652} & \textbf{5.9395} \\
			&	\bfseries Hierarchical  & 0.537515 & 0.5592 &  150.537515 \\
			&	\bfseries K-means & 0.523201& 0.67196& 149.9731 \\
			&	\bfseries BIRCH &0.537515  & 0.5592& 129.5710\\
			
			\hline
			\multirow{5}{*}{\textbf{Iris}} &	\bfseries CA Algorithm  &  \textbf{0.619962} & \textbf{0.5023} & \textbf{441.4913} \\
			&	\bfseries Hierarchical  & 0.600211 & 0.5015 & 387.6531  \\
			&	\bfseries K-means & 0.620466 &0.5021 & 442.8538 \\
			&	\bfseries BIRCH & 0.590375 &0.5027 & 363.9408\\
			
			\hline
			\multirow{5}{*}{\textbf{Buddymove}} &	\bfseries CA Algorithm  &  \textbf{ 0.266941} & \textbf{0.5341} & \textbf{3.101} \\
			&	\bfseries Hierarchical  & 0.244181 & 1.314 &  94.8533 \\
			&	\bfseries K-means & 0.307943 &1.3366 & 119.9733 \\
			&	\bfseries BIRCH & 0.244181 & 1.3140& 94.8533\\
			\hline
			
			\multirow{5}{*}{\textbf{Heart failure clinical records}} &	\bfseries CA Algorithm  &  \textbf{0.581092} & \textbf{1.1893} & \textbf{20.9793} \\
			&	\bfseries Hierarchical  & 0.678929 & 0.4930 &  190.5537 \\
			&	\bfseries K-means & 0.582889 & 0.6319& 329.8704 \\
			&	\bfseries BIRCH & 0.678929 &0.4930 &190.5537 \\
			\hline
			\multirow{5}{*}{\textbf{Cervical Cancer Behavior Risk}} &	\bfseries CA Algorithm  &  \textbf{0.13589} & \textbf{0.7628} & \textbf{1.6125} \\
			&	\bfseries Hierarchical  & 0.27041& 1.4331 &  29.6614 \\
			&	\bfseries K-means & 0.280125 & 1.4204&  32.6010\\
			&	\bfseries BIRCH & 0.27041 & 1.4331& 29.6614\\
			\hline
			\multirow{5}{*}{\textbf{Seeds}} &	\bfseries CA Algorithm  &  \textbf{0.490564} & \textbf{0.7316} & \textbf{322.7543} \\
			&	\bfseries Hierarchical  & 0.51649 & 0.6424 &  308.1367 \\
			&	\bfseries K-means & 0.518287 & 0.6909& 351.1799 \\
			&	\bfseries BIRCH & 0.468223 &0.7496 & 284.5955\\
			\hline
			\multirow{5}{*}{\textbf{Wholesale customers data}} &	\bfseries CA Algorithm  &  \textbf{0.604053 } & \textbf{0.9787} & \textbf{11.9538} \\
			&	\bfseries Hierarchical  & 0.344719 & 1.1973 &  147.4557 \\
			&	\bfseries K-means & 0.511533 & 1.1293&  171.6846\\
			&	\bfseries BIRCH & 0.344719  & 1.1973& 147.4557\\
			\hline
			\multirow{5}{*}{\textbf{StoneFlakes}} &	\bfseries CA Algorithm  &  \textbf{0.481412  } & \textbf{1.9644} & \textbf{3.5399} \\
			&	\bfseries Hierarchical  & 0.413143 &  1.0009 & 46.3611  \\
			&	\bfseries K-means & 0.417812 &1.0150 & 46.5875 \\
			&	\bfseries BIRCH & 0.413143   & 1.0009&46.3611 \\
			\hline
   \multirow{5}{*}{\textbf{Gas Turbine Emission(2015)}} &	\bfseries CA Algorithm  &  \textbf{0.339825 } & \textbf{0.6611} & \textbf{14.8639} \\
			&	\bfseries Hierarchical  & 0.3665 & 1.0908 &  3864.1503 \\
			&	\bfseries K-means & 0.3358  & 1.1796&  4617.4674\\
			&	\bfseries BIRCH &  0.3094    &  1.23756&4203.5965\\
			\hline
	\end{tabular}}
	
\end{table}

\subsection{Implementation and Performance Analysis comparison}
The multistage clustering algorithm we have created can be used in both non-parallel and parallel ways. In the non-parallel approach, rules are sequentially applied for clustering. Conversely, in the parallel implementation, rules can be executed concurrently for the same algorithm utilizing multithreading. Whenever particular rules consistently yield optimal results from the designated list, they are preserved as saved states\cite{manoranjan2023optimized} for our algorithm.\\ \\ \textbf{Saved state:} As our clustering algorithm explores various combinations of the reduced rule set, the system retains the best rule set that produces the optimal clustering score for each dataset. This information is stored in a directory, preserving the best rules as a state for every dataset. If the algorithm has processed a dataset before, it promptly applies the stored rule pair and produces the result. However, if such a rule pair is not found, our algorithm iterates through all possible rules in the best rule set and saves the optimal state for future runs.\\ \\
\textbf{Multi-threading:} The clustering algorithm is multi-threaded, dividing the combinations of rules to be tested into a specified number of threads and executing them concurrently. Multi-threading enhances resource utilization efficiency, as threads share the same memory and data space. This approach proves particularly advantageous for rules that entail longer application times on vertical splits, as running multiple threads concurrently significantly boosts the algorithm's efficiency.

\begin{table}[]
	\centering
	\caption{Performance of our proposed algorithm based on varying split size }\label{Tab:results on various split size}
	\resizebox{0.8\textwidth}{!}{
		\begin{tabular}{|c|c|c|c|c|c|}
			\hline
			{Dataset} & \bfseries Split Size  &  \bfseries Rule 1 & \bfseries Rule 2 & \bfseries Rule 3 & \bfseries Silhouette Score \\
			\hline
			\multirow{5}{*}{\textbf{School District Breakdown}} 
			&	\bfseries 6  &   756019215 & 2018211960  & 1511938590 & 0.7224 \\
			&	\bfseries 7  & 3035673735  & 4030467855 & 3726716205&0.5834  \\
			&	\bfseries 8  & 4031508720 & 756019215 & 2273806215&0.7469\\
			&	\bfseries 9  & 4027544304 & 517140690 & 1259293560& 0.7051\\
			&	\bfseries 10 &  517140690& 252648975 & 267390735 & 0.7469\\
			&	\bfseries 11 &  4030467855 & 755961615 & 757935405 & 0.7469\\
			&	\bfseries 12 & 4041289185 & 1921479288 & 254611245 &0.6716\\
			\hline
			\multirow{5}{*}{\textbf{IPL 2018 Data}} 
			&	\bfseries 6  & 4042321935 & 255652080 & - & 0.4835 \\
			&	\bfseries 7  & 4042321935 & 255652080 & - & 0.4201 \\
			&	\bfseries 8  & 3063191190 & 4031508720 & 4034007024 &0.4997\\
			&	\bfseries 9  & 1263225615 &  252698895& 4041289185 & 0.4275\\
			&	\bfseries 10 & 252698895 & 267390735  &3537972705& 0.4340 \\
			&	\bfseries 11 &   252695055&267390735  &- &0.4340\\
			&	\bfseries 12 &  265482450& 4035440880 & 521018910&0.4889\\
			\hline
			\multirow{5}{*}{\textbf{Iris}} 
			&   \bfseries 6  & 252656880 & 4031508720   & - & 0.6199 \\
			&	\bfseries 7  &  4027576560&  4031508720& -&  0.6032\\
			&	\bfseries 8  & 4027544304 & 4042272240  &- &0.6023\\
			&	\bfseries 9  & 264499440 & 4034007024 & -& 0.6030\\
			&	\bfseries 10 &  252691440 & 264499440  & - & 0.6199\\
			&	\bfseries 11 &  252691440 & 265482450 & - & 0.6199\\
			&	\bfseries 12 &  252691440 &  265482450 & -& 0.6199\\
			\hline
			\multirow{5}{*}{\textbf{Buddymove}} 
			&   \bfseries 6  & 252656880 & 4031508720 & - & 0.2891 \\
			&	\bfseries 7  & 1799965515 & 3726716205 & - &  0.2865\\
			&	\bfseries 8  & 4027544304 & 4030467855 & - & 0.2787\\
			&	\bfseries 9  &  252645360 & 4031508720 & - & 0.2441 \\
			&	\bfseries 10 &  4034007024 & 252691440 & -&0.2891\\
			&	\bfseries 11 & 255652080 & 4027544304 &- &0.2858\\
			&	\bfseries 12 & 4035440880 & 3031741620 & - &0.2664\\
			\hline
	\end{tabular}
 }
	
\end{table}

Now, our algorithm is applied to several standard real-life datasets sourced from (\url{http://archive.ics.uci.edu/ml/index.php}, \url{https://data.world/}), and the scores are determined using benchmark validation indices such as the Silhouette score, Davies-Bouldin index, and Calinski-Harabasz index. Table \ref{tab:Dataset Description} displays the description of the dataset used for clustering using reversible cellular automata. It is important to highlight that higher scores on the Silhouette score and Calinski-Harabasz index signify better clustering, whereas scores closer to zero on the Davies-Bouldin index are indicative of better clustering. In the case of high-dimensional datasets, our algorithm demonstrates good performance. This emphasizes its appropriateness for high-dimensional data and its effectiveness across various datasets. The results are outlined in the Table \ref{Tab:results}. 
The Fig. \ref{fig:Clustering Diagram} illustrates clustering performed by various algorithms on the seed dataset, using perimeter and asymmetry coefficient as the two significant attributes.

Three rules are used for high-dimensional datasets, including Indian Premier League 2018 Batting and Bowling Data (IPL 2018 Data), School District Breakdown Data, and Cervical Cancer Behavior Risk dataset, whereas only two rules are utilized for low-dimensional datasets. Details of the Reversible CA rules used for the clustering of the real-time dataset are given in Table \ref{tab:Dataset Best result Rules}.

We also analyzed the performance of our algorithm for various split sizes of the dataset object using the selected list of rules given in Table \ref{tab:rulelist}. As we vary the split size, both the cell length $n_1$ in Stage 1 and the cell length $n_2$ in Stage 2 also change. With an increase in the split size of $n_1$, the complexity of the rule application increases, necessitating clustering based on $2^{n_1}$ configurations. Consequently, the time complexity of our algorithm depends on the split size. Table \ref{Tab:results on various split size} displays the performance of the datasets as the split size varies. One can observe that our algorithm consistently delivers effective performance across varying split sizes. The implementation of our code is available in \url{https://github.com/kamalikaB/Clustering}.

\textbf{Remark:} Following are some observations:
\begin{itemize}
    \item  The performance of the CA rule slightly differs depending on the split size. So, choosing correct split size is important for effective clustering.
    \item In high-dimensional datasets, selecting rules that will create fewer cycles with the real dataset in Stage 1 reduces the merged object data size, enabling the application of a new rule in Stage 2  without the need for a recursive call to Stage 1. This ultimately saves computational time.
    \item If getting the best clusters is not a strict requirement, user can run the algorithms for desired number of trials and take the best result of those trials stored in the directory for saved states.
\end{itemize}

\section{Conclusion}

This paper introduces a novel three-stage clustering algorithm designed specifically for high-dimensional datasets, applicable to various fields including healthcare, chemical research, agriculture, and more. The methodology involves the segmentation of frequency-based encoded dataset objects, utilizing the maximum feasible cell size. The integration of a carefully selected set of CA rules significantly improves the runtime efficiency of the clustering algorithm. Subsequently, we evaluated the clustering quality through various performance analysis measures and conducted a comparative analysis with several state-of-the-art clustering algorithms. The results indicate that our proposed algorithm outperforms existing methods for many datasets, showcasing its effectiveness. However, our proposed algorithm exhibits lower performance for certain datasets according to evaluation metrics such as the Calinski-Harabasz index. Addressing this issue could be considered as a direction for future work on the proposed model.

\section*{Acknowledgment}
The authors gratefully acknowledge the contribution of Mr. Abhishek, S., Mr. Mohammed Dharwish, Mr. Amit Das, Mr. Viswonathan Manoranjan, Ms. G. Sneha Rao and Mr. Subramanian V. V. for their contribution to build up the basis of this work and also to the last three for their efficient coding, available in GitHub repository: \url{https://github.com/Viswonathan06/Reversible-Cellular-Automata-Clustering}, which has been reused in the paper for comparison and testing.

\section*{Declarations} 

\textbf{Ethical approval:} This is not applicable. \\

\noindent\textbf{Competing interests:} The authors have no relevant financial or nonfinancial interests to disclose.\\

\noindent\textbf{Author contributions:} \\Baby C J - Validation, Formal analysis, Investigation, Writing - Original Draft, Software, Visualization, Data Curation.\\
Kamalika Bhattacharjee- Conceptualization, Methodology, Writing - Review and Editing, Supervision, Funding acquisition.\\

\noindent\textbf{Funding:} 
This work is partially supported by Start-up Research Grant (File number: SRG/2022/002098), SERB, Department of Science \& Technology, Government of India, and NIT, Tiruchirappalli SEED Grant. \\

\noindent\textbf{Data availibility:} No Data associated in the manuscript. The implementation of our code is available in \url{https://github.com/kamalikaB/Clustering}.

\bibliographystyle{ws-acs}
\bibliography{References_thesis}

\end{document}